# Booster 6-GeV Study

Xi Yang, Charles M. Ankenbrandt, William A. Pellico, James Lackey, and Rene Padilla
*Fermi National Accelerator Laboratory*
Box 500, Batavia IL 60510
J. Norem
*Argonne National Laboratory*
9700 S. Cass Ave, Argonne, IL, 60439

**Abstract**

Since a wider aperture has been obtained along the Booster beam line, this opens the opportunity for Booster running a higher intensity beam than ever before. Sooner or later, the available RF accelerating voltage will become a new limit for the beam intensity. Either by increasing the RFSUM or by reducing the accelerating rate can achieve the similar goal. The motivation for the 6-GeV study is to gain the relative accelerating voltage via a slower acceleration.

## Introduction

After the Booster-lattice distortion caused by the edge focusing of dogleg magnets in extraction sections of long 3 and long 13 was largely removed via the rearrangement of spacing the magnets out, one expects that a much wider aperture will be available for the acceleration of higher intensity Booster beams. Also, since the collimator has been effectively reducing the radiation loss (RL) after its installation, the chance that the RL becomes the limit for the highest beam intensity (HBI) is very small. Sooner or later, the RF accelerating voltage (RFSUM) ($V_{RF}$) from all the Booster RF stations,[1] especially at the transition crossing (TC), will become a crucial factor for determining the HBI. The reason is because the effective accelerating voltage ($V_{eff}$) is equal to the sum of two parts ($V_a$ and $V_L$), and it is the accelerating voltage seen by the beam, as shown by eq.1.

$$\begin{aligned} V_{eff} &= V_a + V_L \\ &= V_{RF} \times \sin(\varphi_s). \end{aligned} \tag{1}$$

Here, $\varphi_s$ is the synchronous phase, which is the phase difference between the RFSUM and the centroid of the circulating beam bunch (CB).[2]

$V_a$ is the accelerating voltage required by the rate of change of the Booster magnetic field (d$B$/d$t$) in a cycle, and is independent of the beam intensity. $V_L$ is the beam energy loss, which is caused by the real impedance of the ring, and is dependent upon the beam current.[3] The higher the beam current is, the higher the beam energy loss is. It is clear that $V_{eff}$ is limited by $V_{RF}$. The HBI is achieved whenever $V_L$ reaches the difference between $V_{RF}$ and $V_a$ in a cycle. From past experience in the normal 8-GeV operation, this usually happens at the TC when '$V_{RF}$ -$V_a$' reaches its minimum while $V_L$ reaches its peak value. This is because d$B$/d$t$ reaches its peak at the TC, while the peak current of the beam reaches its maximum as a result of the shortest bunch length.

Either by increasing the RFSUM or by reducing the accelerating rate, the allowable energy loss, which is the difference between $V_{RF}$ and $V_a$, will be increased, and also the HBI should be increased. It is obvious that the maximum number of RF stations, which can be installed in Booster, together with the output of each station, constrain the upper limit of $V_{RF}$. Another way of increasing '$V_{RF}$ -$V_a$' is to reduce $V_a$ via a slower acceleration, and this is the motivation for the 6-GeV study.

## Experimental Result and Analysis

All the relevant Booster ramps, which include the magnet ramp, the RF frequency ramp, the bias ramp, etc., were rescaled to the 6-GeV acceleration. Also, some routine tunings, which include the injection tuning, RF curve tuning, quad ramp tuning, etc., had been performed before Booster reached a standard running condition.

Afterwards, for the purpose of finding the relationship between the phase jump at the TC and the beam intensity, the synchronous phase detector was used to measure the $\varphi_s$ right before the TC and right after the TC, and their difference gives the phase jump ($\Delta\varphi_s$). Here, the Linac beam was injected at 0 ms, and the TC gate was set at 20.7 ms.

The same measurement was repeated for several extracted beam intensities and also at two different RFSUM curves, RFSUM 1 and RFSUM 2. The results were compared to the data which were taken at the 8-GeV operation.[5] RFSUM 1, RFSUM 2, and RFSUM for the 8-GeV acceleration are shown as the black, red, and green curves in Fig. 1(a) respectively. Their corresponding $\Delta\varphi_s$ *vs*. the extracted beam intensity and linear-fit results are shown in Fig. 1(b). The operational region with extracted beam intensities, which are less than or equal to $6\times10^{12}$ protons, is indicated by the red rectangular in Fig. 1(b).

The accelerating voltages required by d*B*/d*t* for the 6-GeV acceleration and 8-GeV acceleration are shown as the black and red curves in Fig. 2(a) respectively.[2] The synchronous phase and RFSUM were measured at extracted beam intensities of $4.1\times10^{12}$ protons, $3.6\times10^{12}$ protons, $1.9\times10^{12}$ protons, $0.35\times10^{12}$ protons when Booster was operated at RFSUM 1, and they are shown in Figs. 2(b) and 2(c). And their effective accelerating voltages ($V_{eff}$) per beam turn (BT), which were calculated using eq.1, are shown in Fig. 2(d). The black, red, green, and blue curves in each plot represent the results for the several intensities from highest to lowest respectively. The magenta curve in Fig. 2(d) is the same with the black curve in Fig. 2(a). The intensity dependent part (IDP) of the $V_{eff}$, which is equivalent to the IDP of the $V_L$, can be estimated from differences between the black and blue curves, the red and blue curves, the green and blue curves in Fig. 2(d). Furthermore, by taking the difference, the error coming from the offset of signals $V_{RF}$ and $\varphi_s$ can be minimized, and their differences are shown as the black, red, and green curves in Fig. 2(e) respectively. The same procedure was applied to the situation when Booster was operated at RFSUM 2. The $V_{eff}$ at extracted beam intensities of $4.7\times10^{12}$ protons, $3.8\times10^{12}$ protons, $2.1\times10^{12}$ protons, and $0.39\times10^{12}$ protons are shown as the black, red, green and blue curves in Fig. 2(f) respectively, and the magenta curve is the same with the one in Fig. 2(d). Differences between the black and blue curves, the red and blue curves, the green and blue curves in Fig. 2(f) are shown as the black, red, and green curves in Fig. 2(g). The IDP of $V_L$ in the RFSUM 1 situation, as shown in Fig. 2(d), are similar to those in the RFSUM 2 case, as shown in Fig. 2(g). This is what one expects, since there is an approximately linear relationship between the IDP of $V_L$ and the beam intensity. Unless when the CB gets too short or the shape of the bunch is no longer smooth, high frequency components of the beam current couldn't be neglected any more.

The $V_{eff}$ difference between operations of RFSUM 1 and RFSUM 2 for the similar extracted beam intensity can give us some ideas of how to choose the optimal RFSUM curve in a cycle. The smaller the $V_{eff}$ is, the less the effective $V_L$ is. The $V_{eff}$ differences between RFSUM 1 and RFSUM 2 at extracted beam intensities of $3.7\times10^{12}$ protons and $0.36\times10^{12}$ protons are shown as the black and red curves in Fig. 3(a). It is clear that RFSUM 1 is better than RFSUM 2 before the TC and is worse after the TC. The green curve with arrows in Fig. 3(b) indicates what the likely optimal RFSUM curve might be.

All data were taken under similar conditions with the extracted beam intensity about $5.1\times10^{12}$ protons when Booster was operated at RFSUM 2 of 6 GeV. The method for determining the lower limit for RFSUM is to reduce it to the value at which the beam loss starts.[6] In the experiment, the RFSUM limit was measured at 3.5 ms, 6.5 ms, 9.5 ms, 12.5 ms, 15.5 ms, 18.5 ms, 21.5 ms, 24.5 ms, and 27.5 ms separately. The RFSUM, the RFSUM limit, the $V_{eff}$, and the $V_a$ are shown as the black, red, green, and blue curves respectively in Fig. 4(a). The data at the 8-GeV operation for the similar beam intensity are shown in Fig. 4(b) for the purpose of comparison.[6] It is clear that in the 6-GeV acceleration, there are more RF voltages that are available for the compensation of the beam energy loss, especially for the high intensity beam.

RF bucket reduction for acquiring information about the particle distribution in longitudinal phase space was applied for the extracted beam intensity of $4.3\times10^{12}$ protons.[7] We chose eight points, 3.5 ms, 6.5 ms, 9.5 ms, 12.5 ms, 18.5 ms, 21.5 ms, 24.5 ms, and 27.5 ms in a Booster cycle to do the measurement. At 3.5 ms, RFSUM was reduced to nine different values, as shown in Fig. 5(a), and their corresponding charges, which were left in the RF bucket, were recorded as Fig. 5(b). Also, the synchronous phase was recorded at three different RFSUM values, and they were used to find the relationship between the RFSUM value and synchronous phase via the $2^{nd}$ order polynomial fit, and the result is shown in Fig. 5(c). The RFSUM and the synchronous phase *vs*. charge are shown as the black and blue curves respectively in Fig. 5(d). Eq.2 is used to calculate the bucket area ($A$) in unit of eV×s.[8]

$$A = \frac{16 \cdot \beta}{2 \cdot \pi \cdot f_{rf}} \cdot \sqrt{\frac{e \cdot V_{RF} \cdot E_s}{2 \cdot \pi \cdot h \cdot |\eta|}} \cdot \alpha(\varphi_s). \tag{2}$$

Here, $\beta$ is the Lorentz's relativistic factor, $f_{rf}$ is the RF frequency, $\eta$ is the phase slip factor, $E_s$ is the kinetic energy, $h$ is the harmonic number of the rf frequency, $e$ is the electron charge, and $\alpha(\varphi_s)$ is the ratio of the bucket area with $\varphi_s$ relative to the stationary bucket area (either $\varphi_s$=0°, or $\varphi_s$=180°).[8] The charge *vs*. bucket area is shown in Fig. 5(e). The differentiation of the charge over the bucket area gives the charge density in the bucket area, as shown in Fig. 5(f). The same procedure was applied for getting the relationship between the charge density and the bucket area at the rest time of the cycle, and the results are shown in Figs. 5(g)-5(m). The charge-density fluctuation inside the bucket area appeared strongly right after the TC, as shown in Fig. 5(k), lasted till 24.5 ms, and was smoothed at 27.5 ms.

## Comment

The effective accelerating voltage was reduced at the 6-GeV acceleration since the accelerating rate was reduced in the 6-GeV cycle compared to the 8-GeV cycle. However, the intensity-dependent part of the energy loss, or called the AC component of the energy loss, does not vary significantly from 8-GeV acceleration to 6-GeV acceleration and from one RFSUM curve to another, as shown in Fig. 2(e) and Fig. 2(g). The intensity-independent part of the energy loss, or called the DC component of the energy loss, does depend upon the RFSUM curve, as shown in Fig. 3(a). So optimizing the RFSUM curve still will be helpful for reducing the DC component of the energy loss. The strong charge density fluctuation in the bucket right after the TC, which is likely caused by the mismatch of the beam in the bucket before and after the TC, might be the intrinsic property of the Booster RF system, including LLRF and HLRF, and this only can be fixed by upgrading the RF system or commissioning the $\gamma_t$ jump system for the purpose of making the TC faster.

## Acknowledgement

Authors give special thanks to Craig Drennan for the instruction of how to use the new paraphrase module.

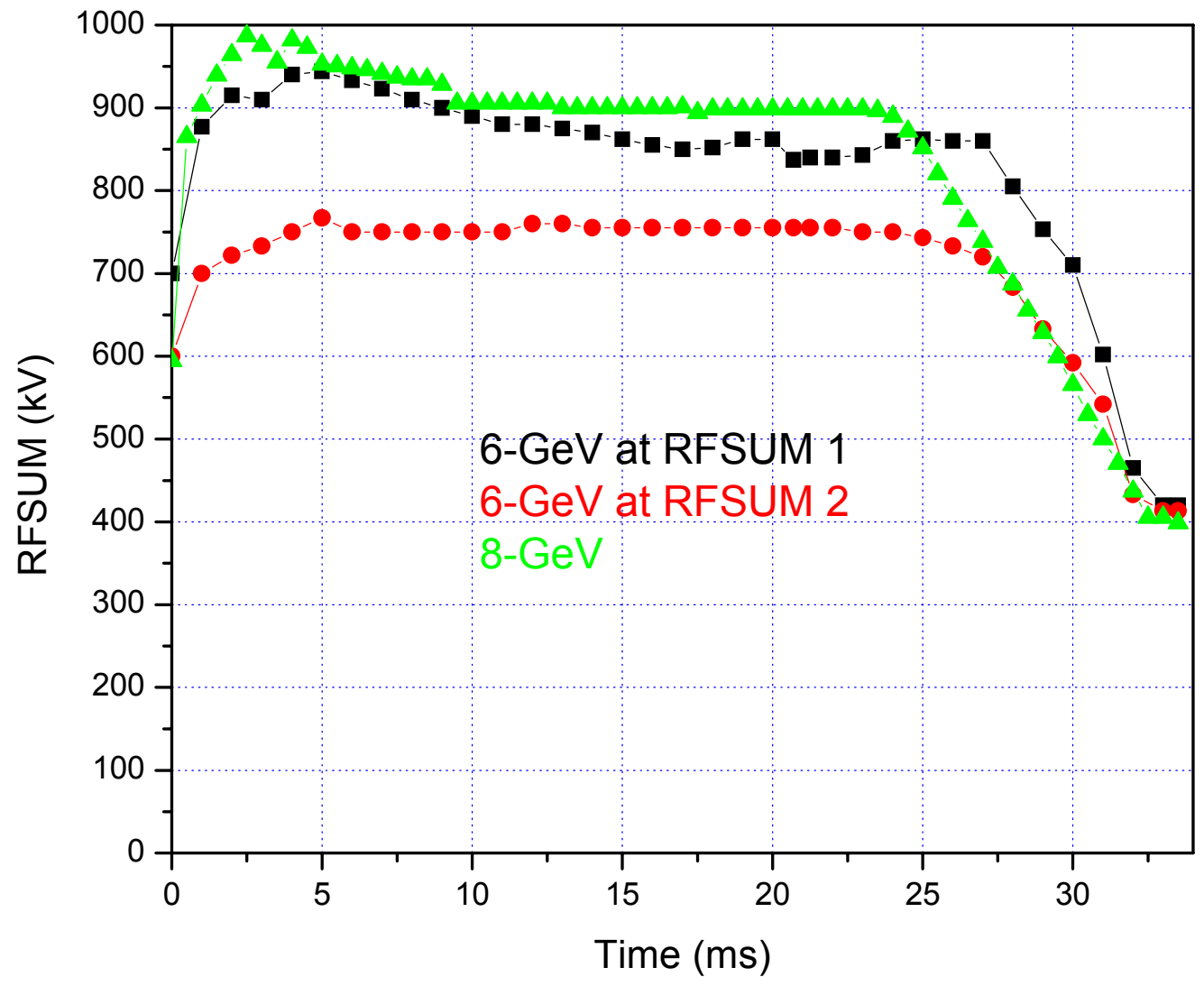

Fig. 1(a)

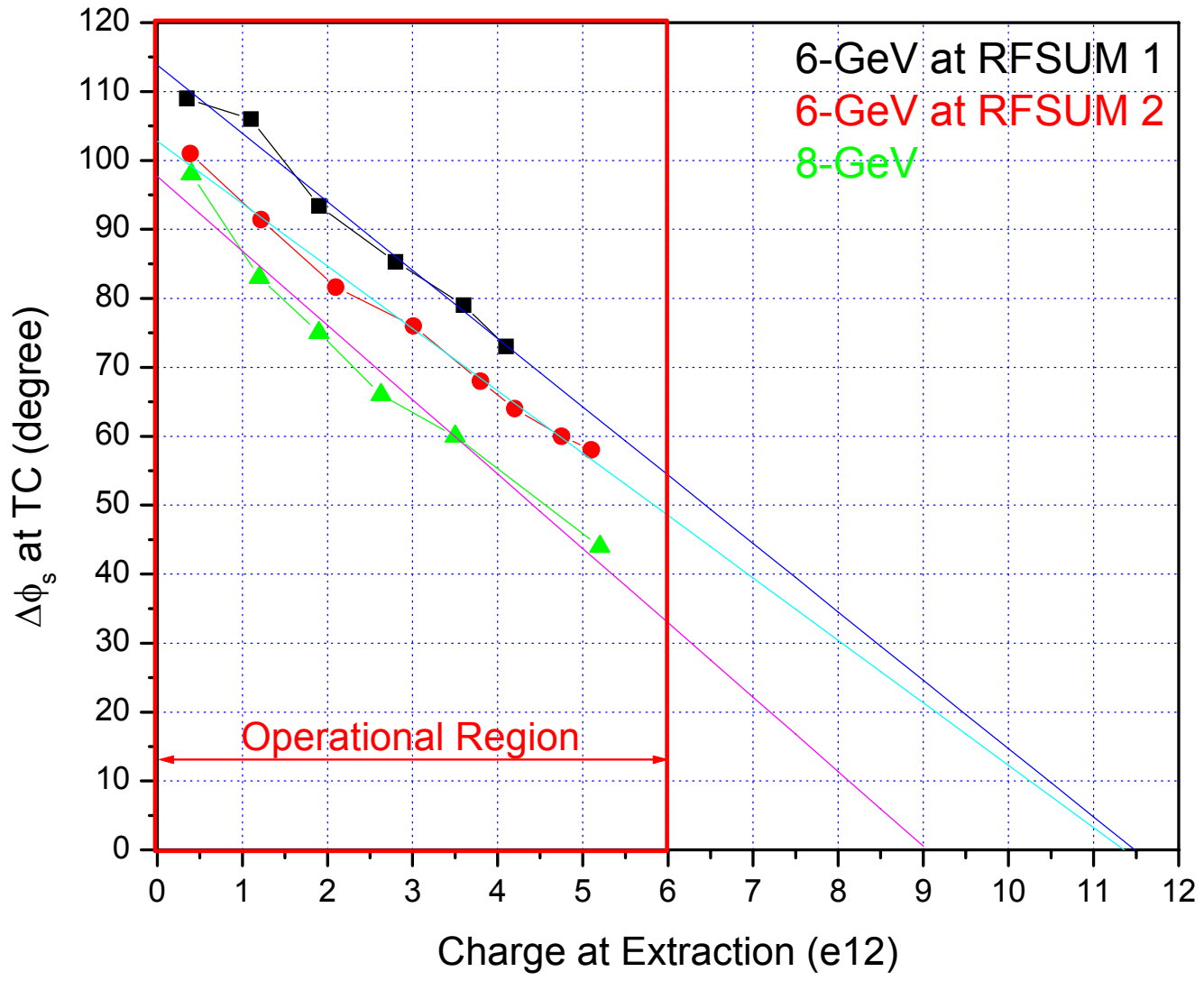

Fig. 1(b)

Fig. 1(a) RFSUM 1 and RFSUM 2 for the 6-GeV acceleration and RFSUM for the 8-GeV acceleration are shown as the black, red, and green curves respectively.

Fig. 1(b) their corresponding $\Delta\varphi_s$ *vs*. the extracted beam intensity of Fig. 1(a).

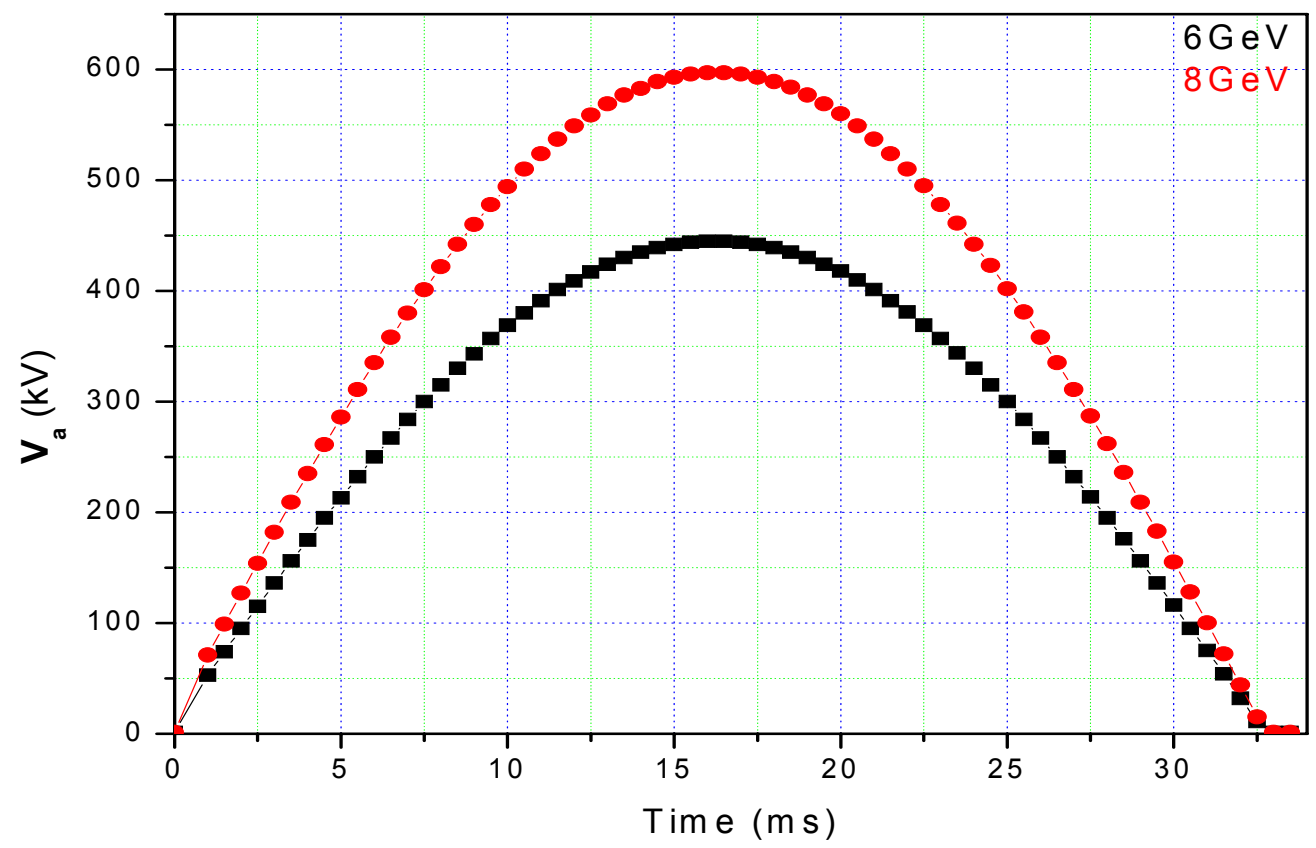

Fig. 2(a)

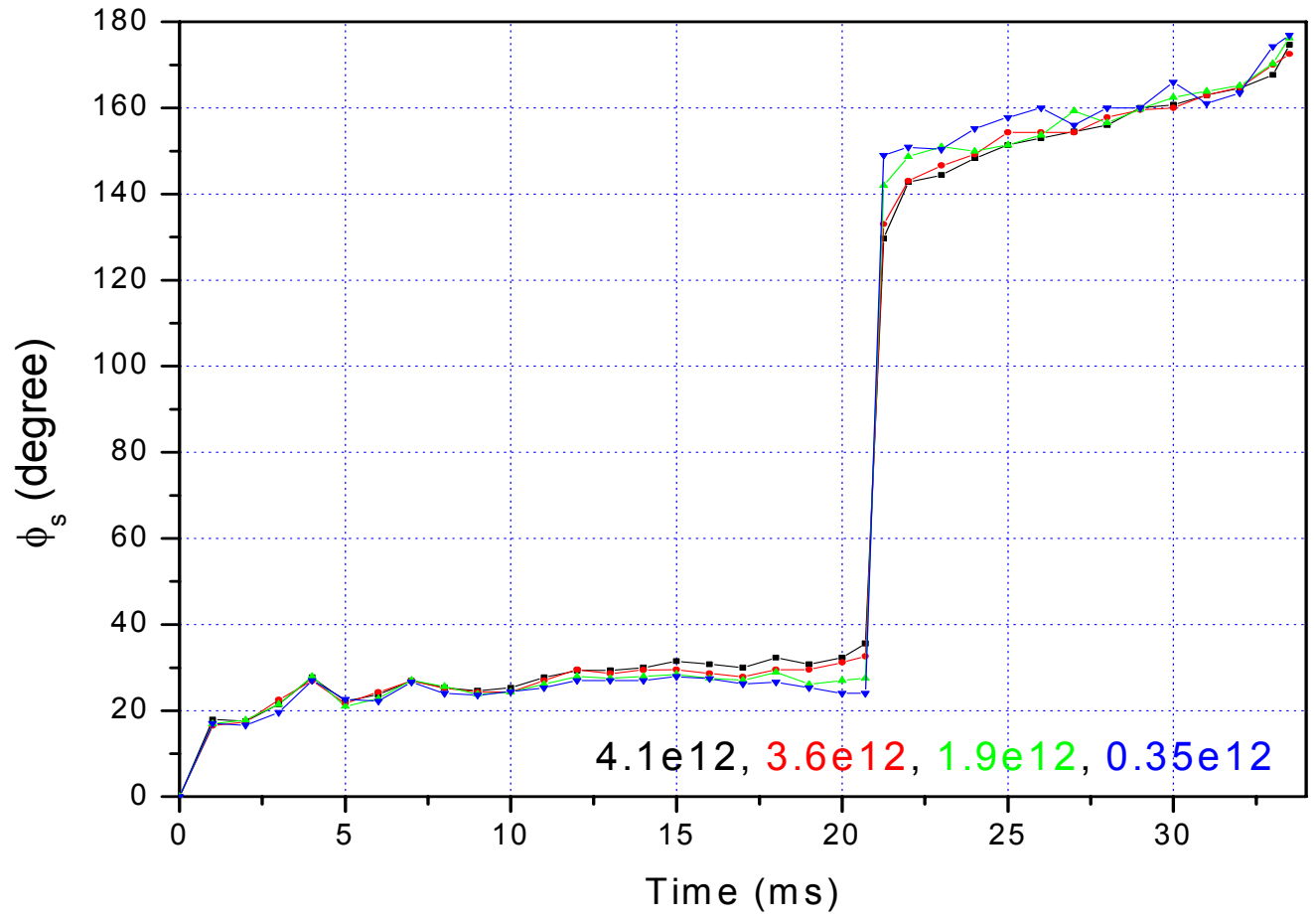

Fig. 2(b)

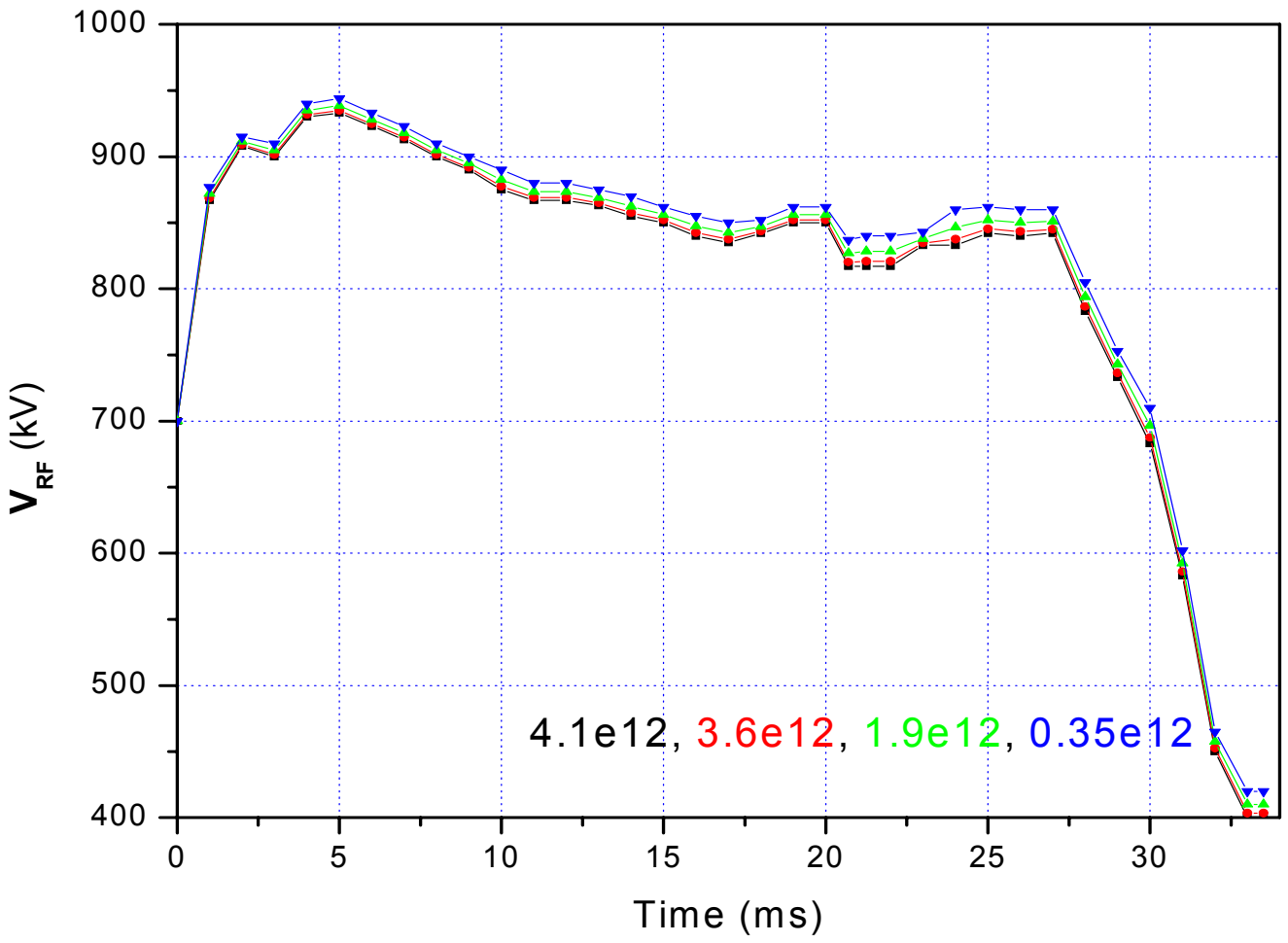

Fig. 2(c)

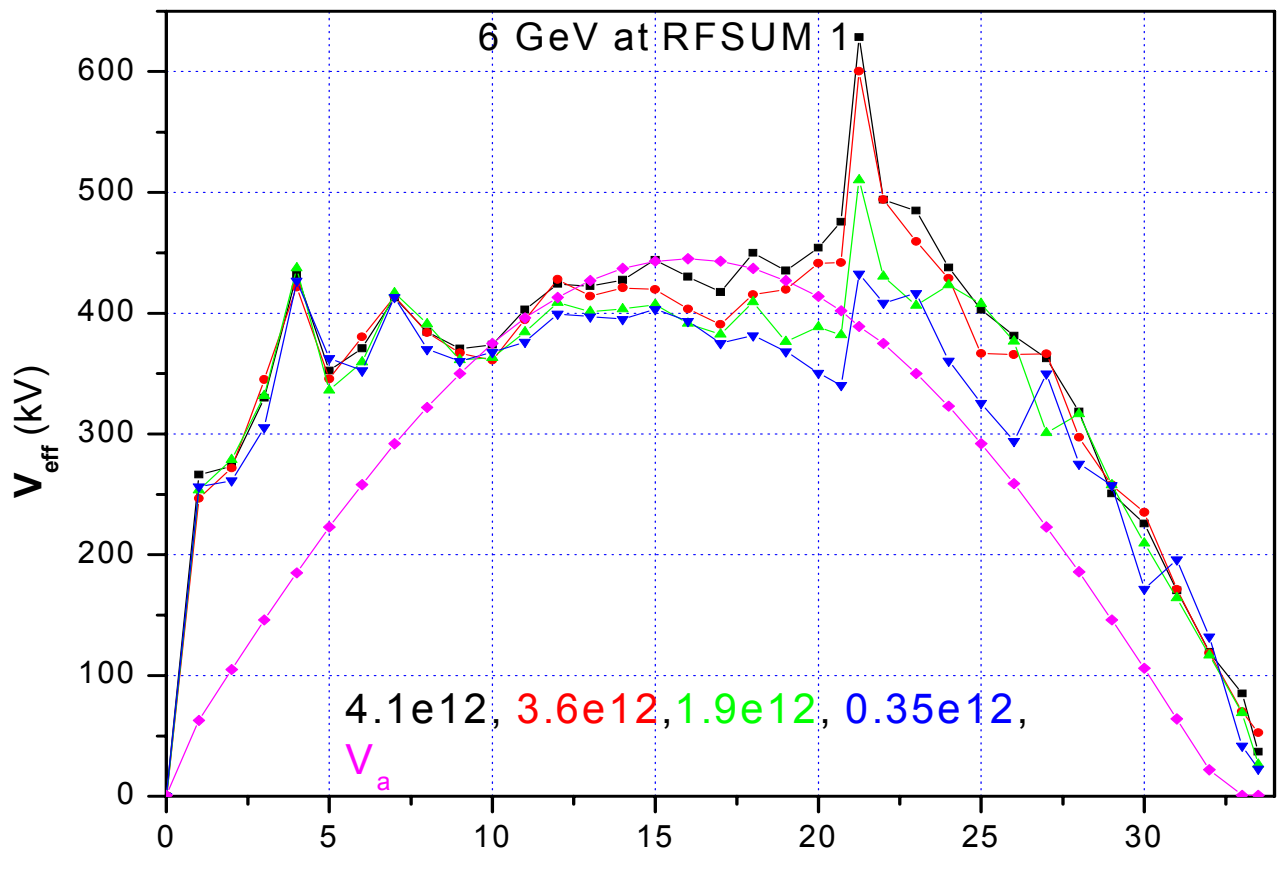

Fig. 2(d)

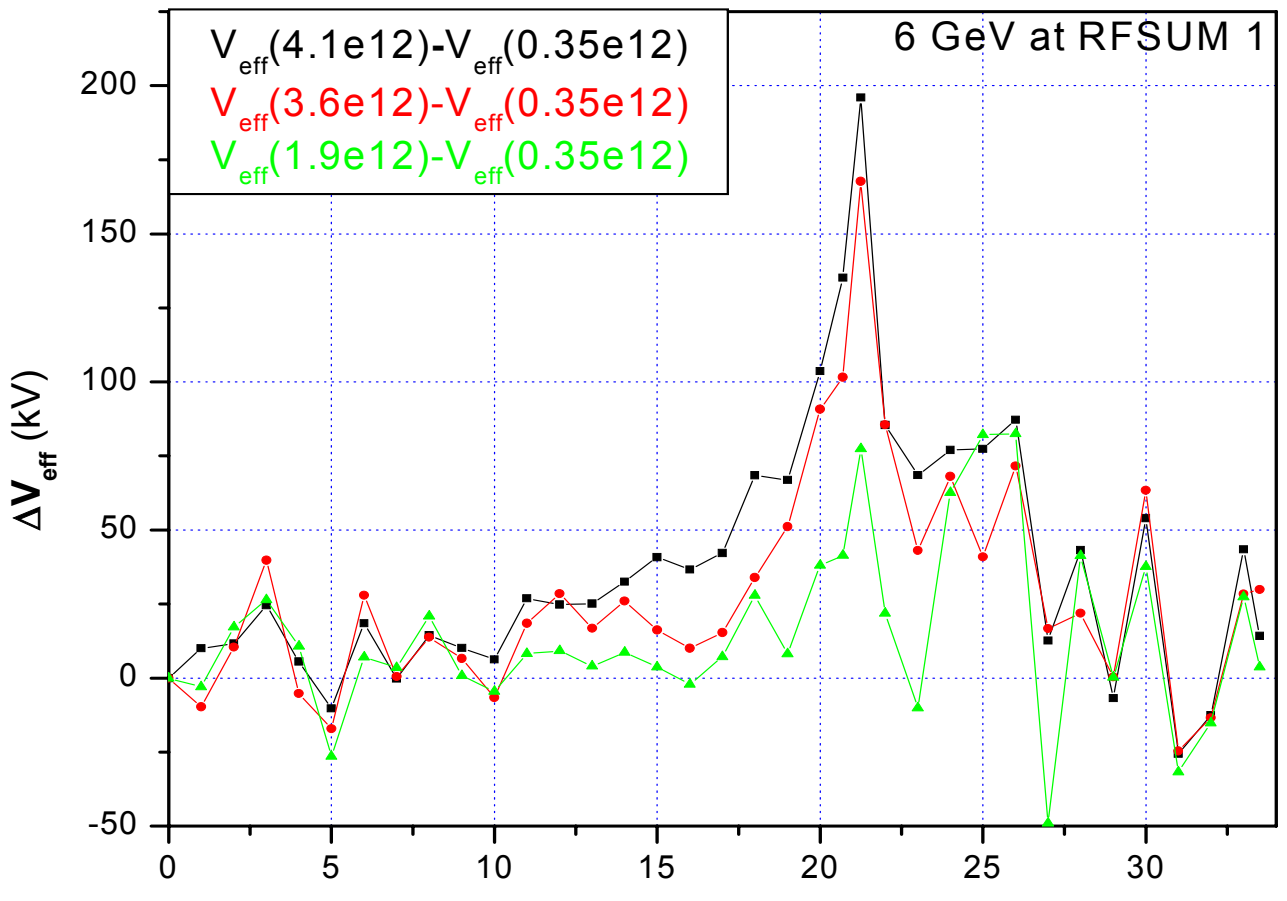

Fig. 2(e)

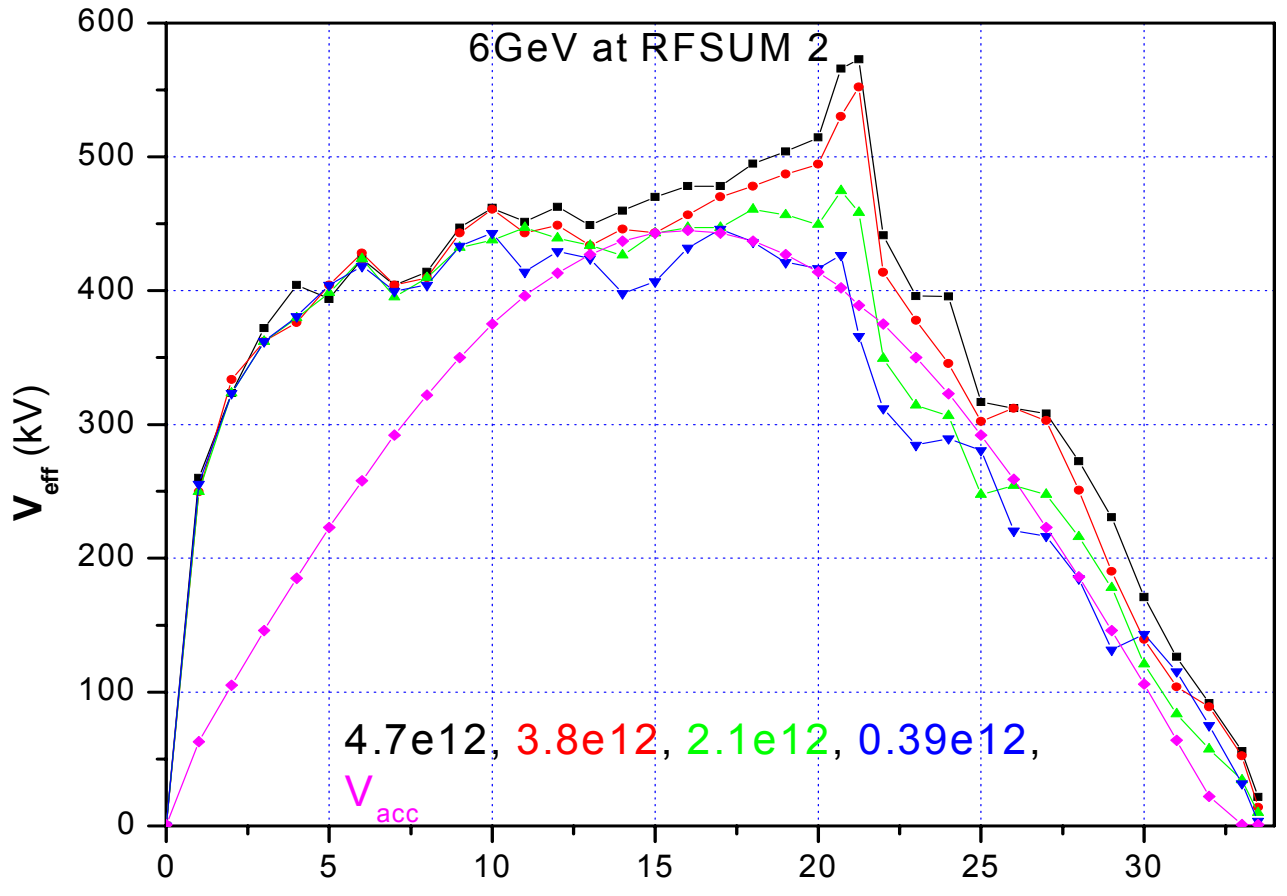

Fig. 2(f)

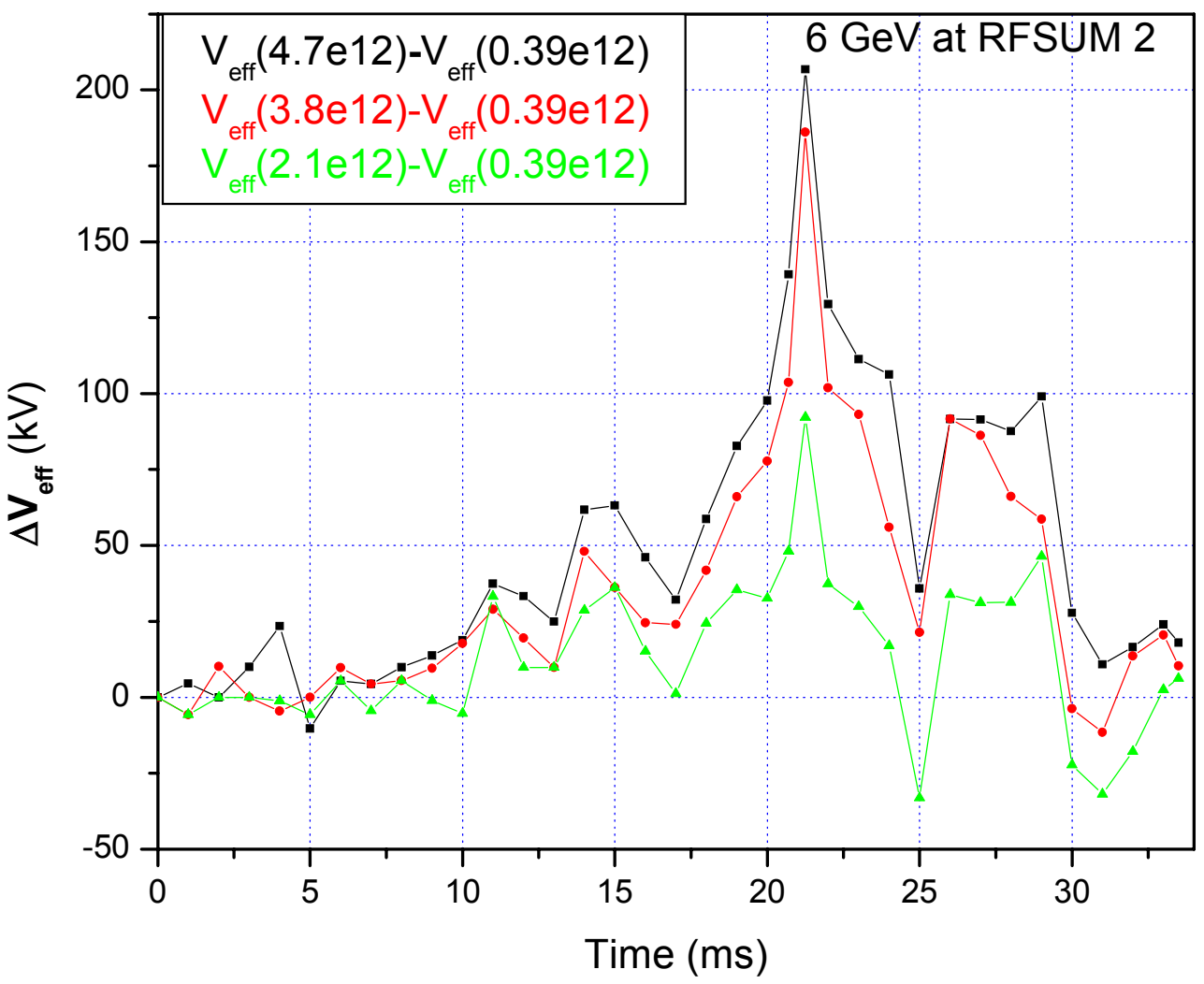

Fig. 2(g)

Fig. 2(a) the accelerating voltages required by d*B*/d*t* for the 6-GeV acceleration and 8-GeV acceleration are shown as the black and red curves respectively.

Fig. 2(b) the synchronous phase *vs*. time in a cycle taken at extracted beam intensities of $4.1\times10^{12}$ protons (the black curve), $3.6\times10^{12}$ protons (the red curve), $1.9\times10^{12}$ protons (the green curve), $0.35\times10^{12}$ protons (the blue curve).

Fig. 2(c) RFSUM *vs*. time.

Fig. 2(d) the effective accelerating voltage $V_{eff}$ *vs*. time.

Fig. 2(e) the $V_{eff}$ differences between the black and blue curves, the red and blue curves, the green and blue curves in Fig. 2(d) are shown as the black, red, and green curves respectively.

Fig. 2(f) $V_{eff}$ at extracted beam intensities of $4.7\times10^{12}$ protons, $3.8\times10^{12}$ protons, $2.1\times10^{12}$ protons, and $0.39\times10^{12}$ protons are shown as the black, red, green and blue curves respectively when Booster operated at RFSUM 2.

Fig. 2(g) the $V_{eff}$ differences between the black and blue curves, the red and blue curves, the green and blue curves in Fig. 2(f) are shown as the black, red, and green curves respectively.

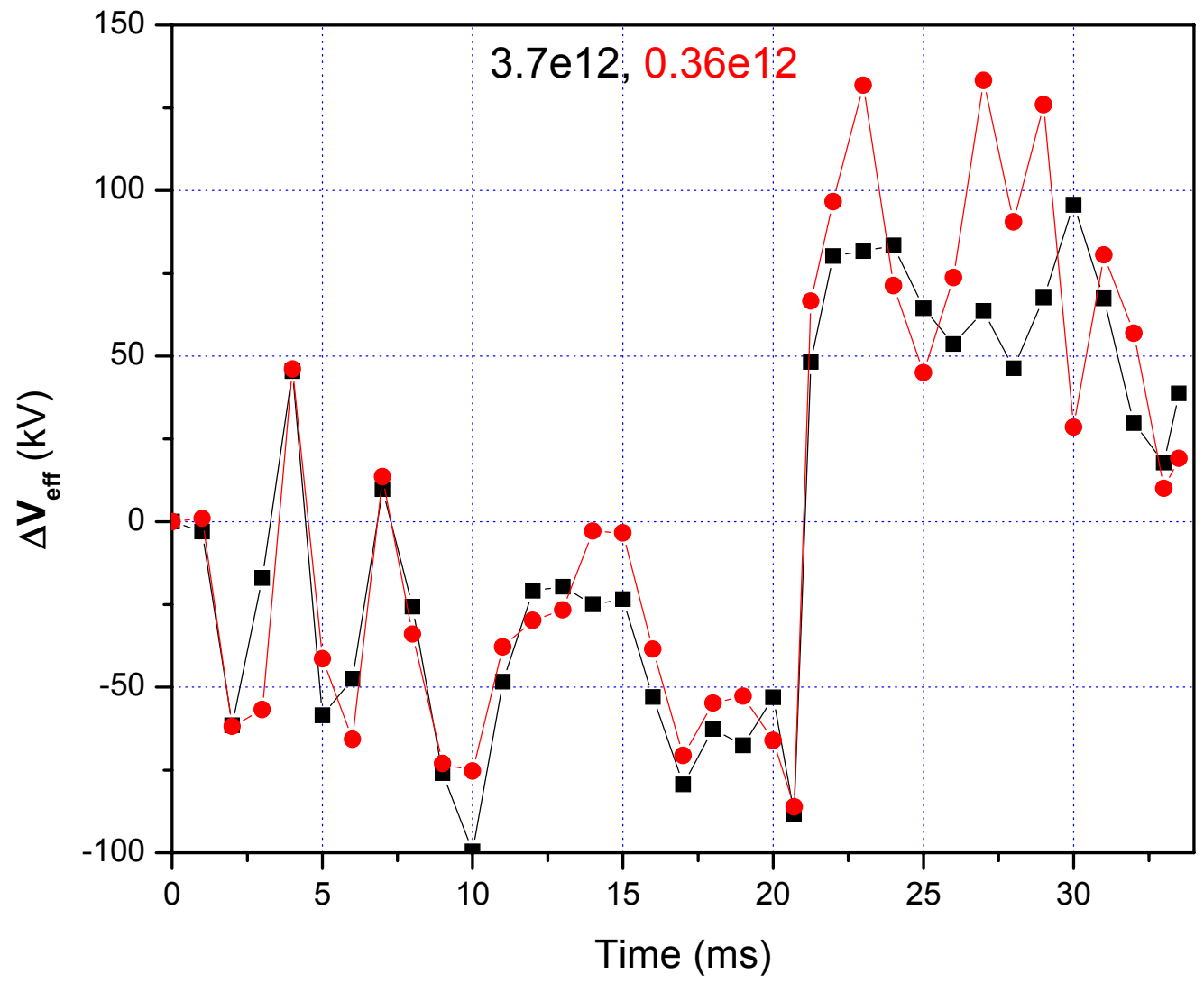

Fig. 3(a)

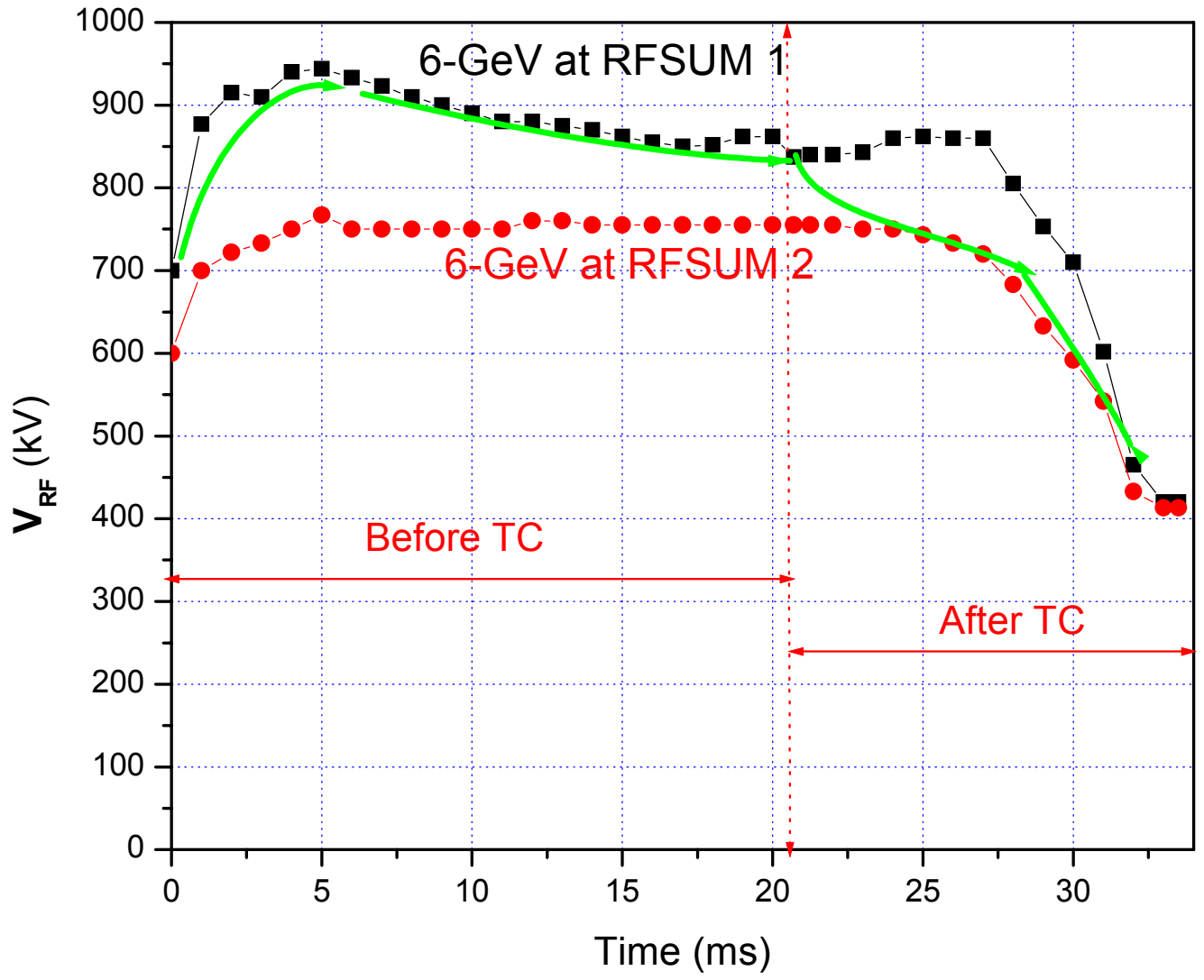

Fig. 3(b)

Fig. 3(a) the $V_{eff}$ differences between RFSUM 1 and RFSUM 2 at extracted beam intensities of $3.7\times10^{12}$ protons and $0.36\times10^{12}$ protons are shown as the black and red curves respectively.

Fig. 3(b) RFSUM 1 and RFSUM 2 are shown as the black and red curves separately. The green curve indicates where the optimal RFSUM curve might be.

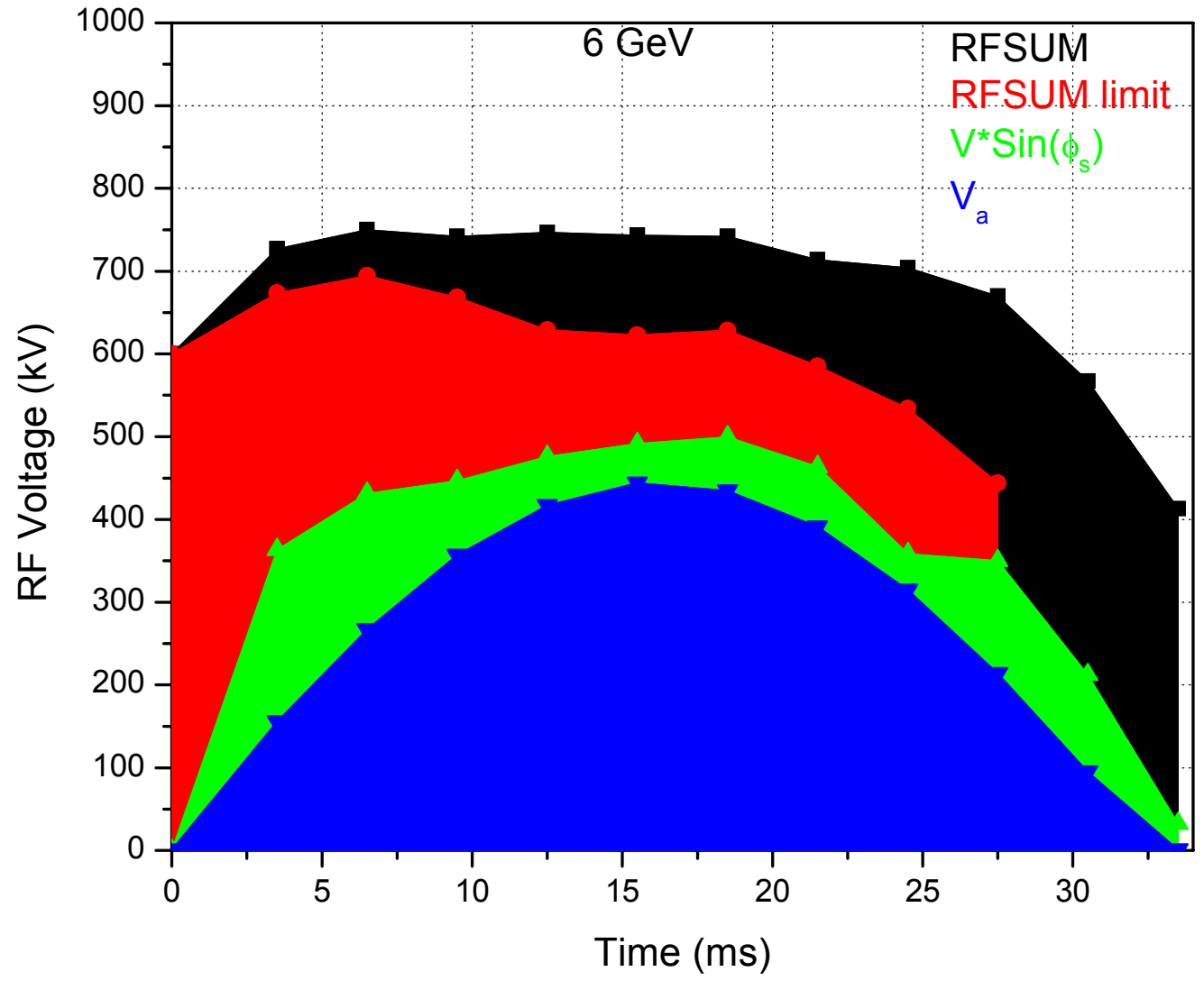

Fig. 4(a)

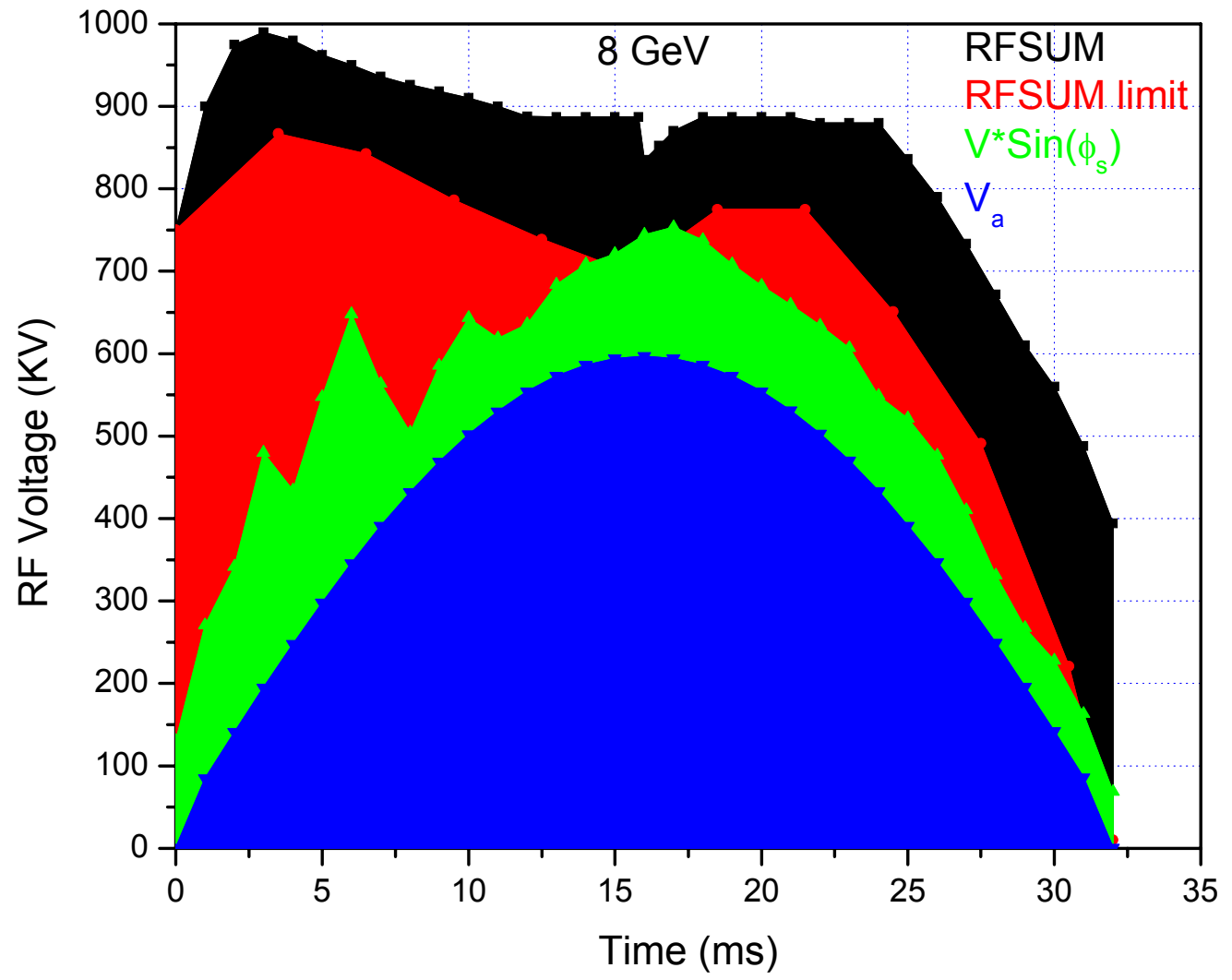

Fig. 4(b)

Fig. 4(a) the black curve represents RFSUM in a Booster cycle, the red curve represents the lower limit for RFSUM, the green curve represents the effective accelerating voltage, and the blue curve represents the accelerating voltage required by the magnet ramp, all data were taken under closely similar conditions at the extracted beam intensity of $5.1\times10^{12}$ protons when Booster was operated at the 6 GeV.

Fig. 4(b) the same situation with Fig. 4(a), except the data were taken when Booster was operated at the 8 GeV.

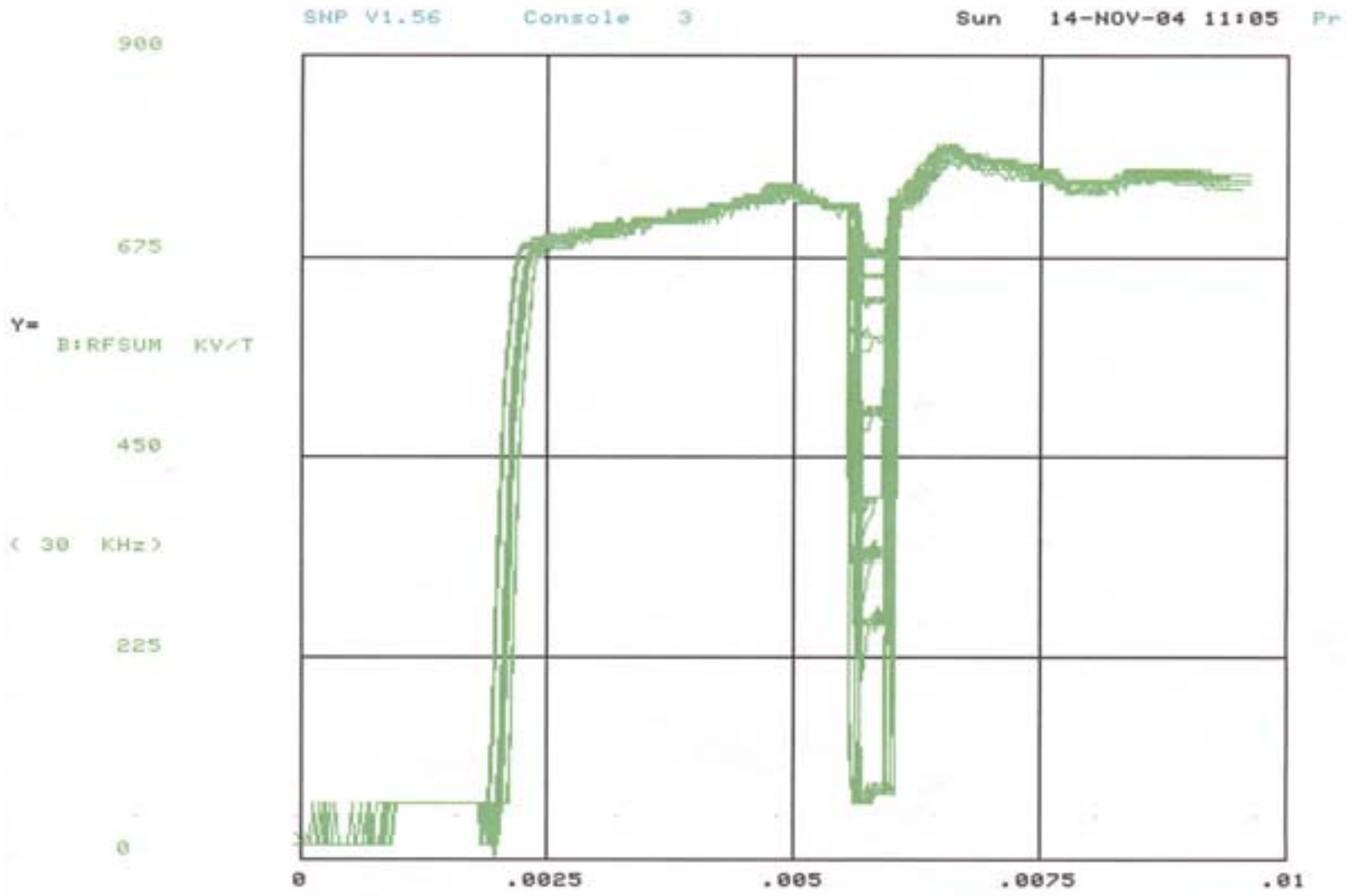

Fig. 5(a)

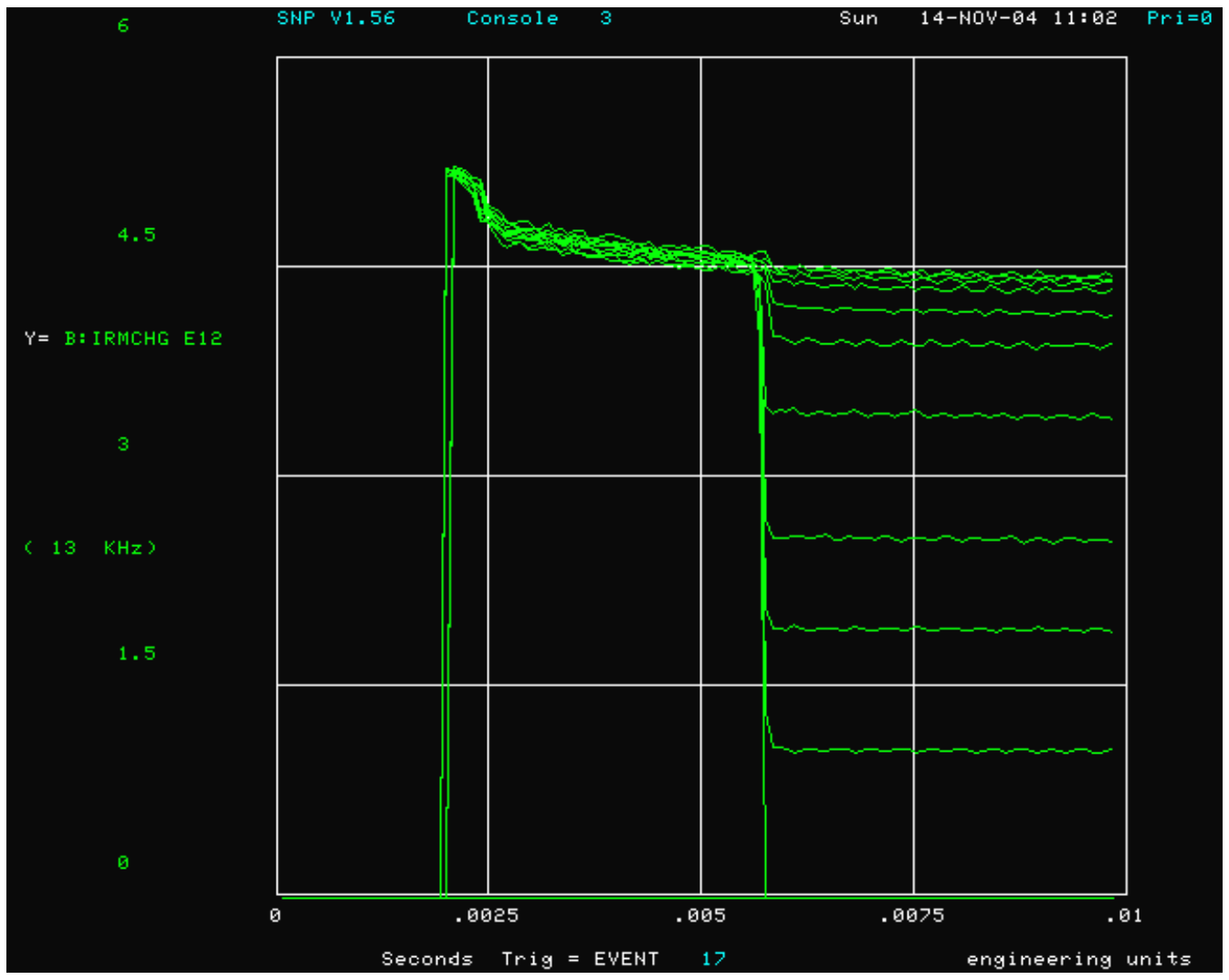

Fig. 5(b)

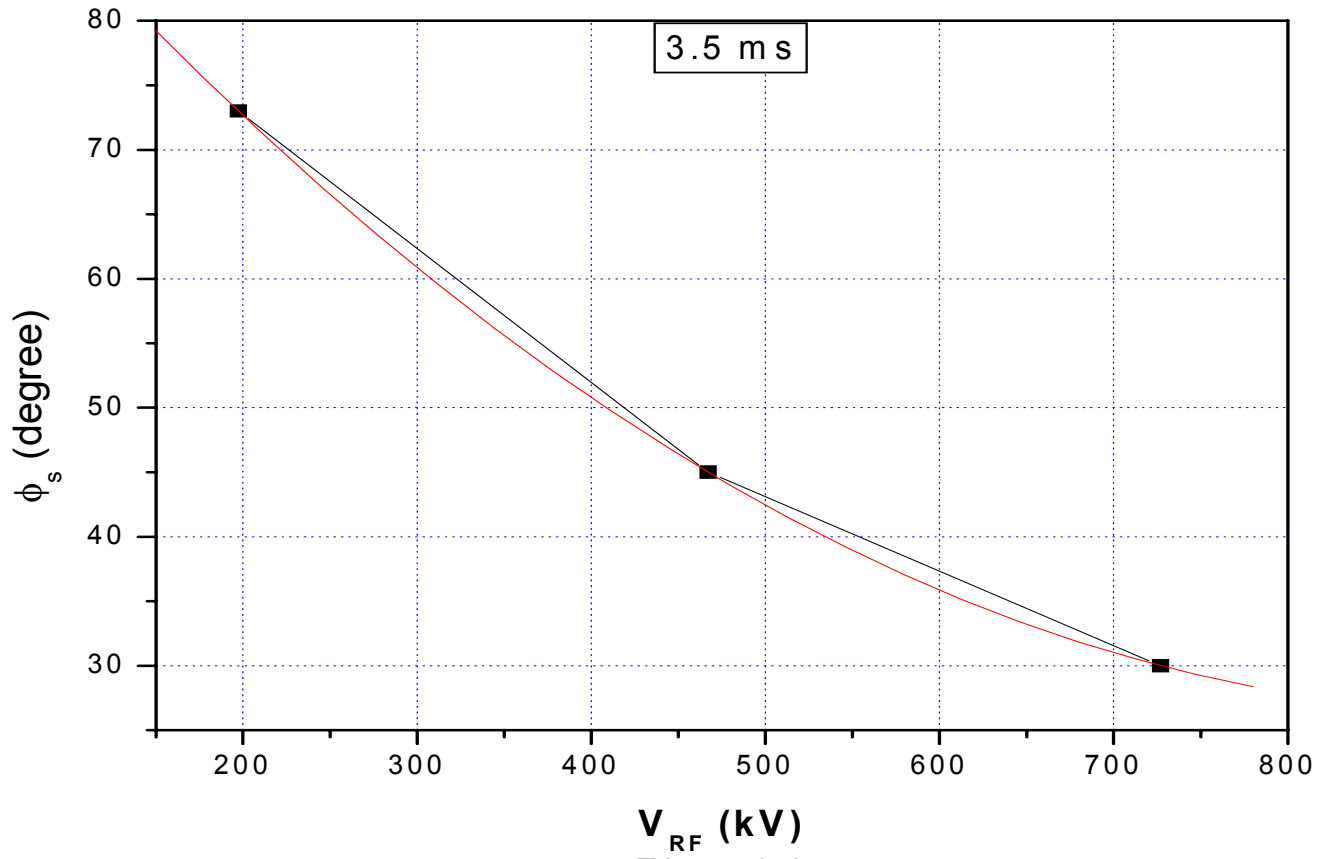

Fig. 5(c)

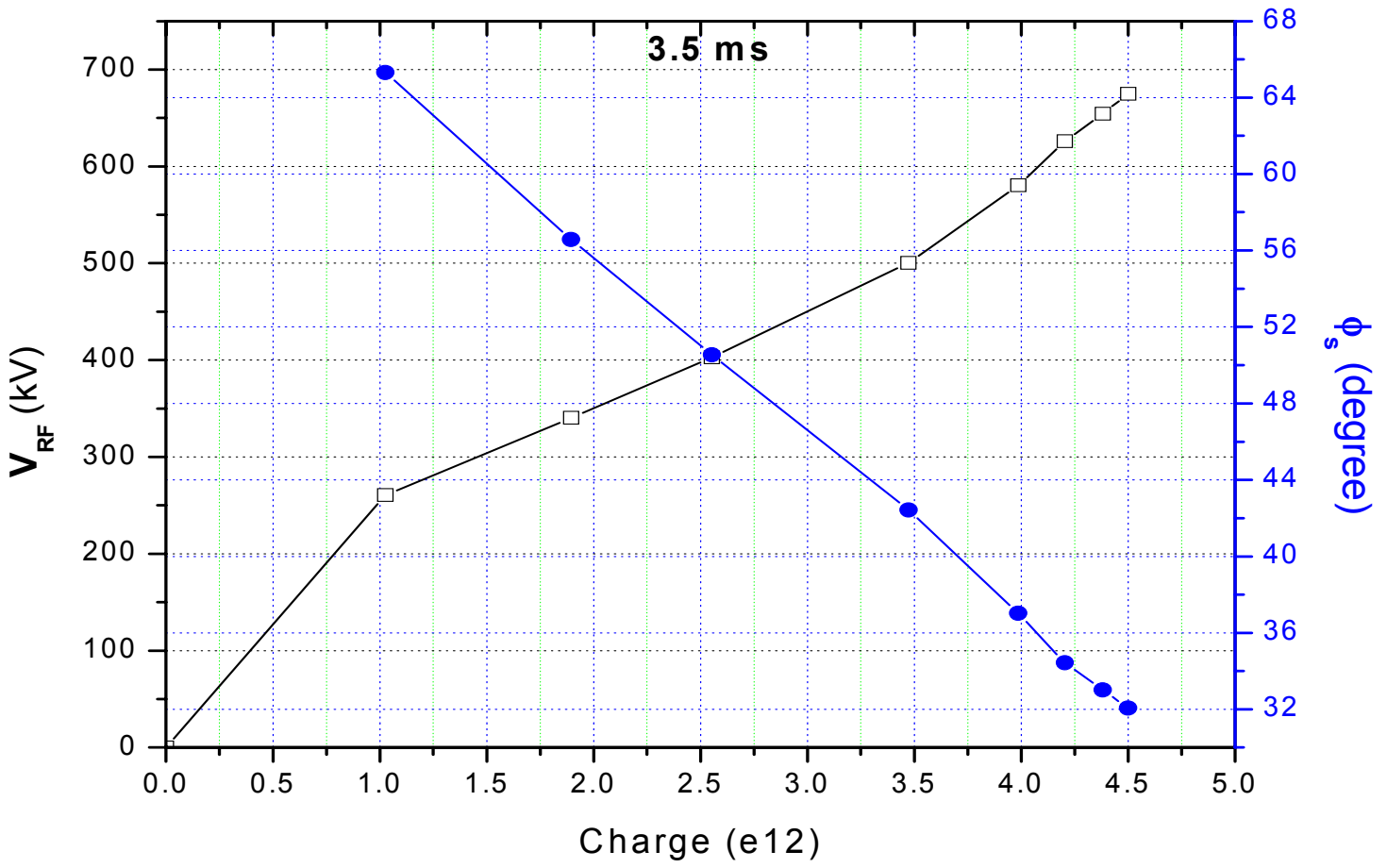

Fig. 5(d)

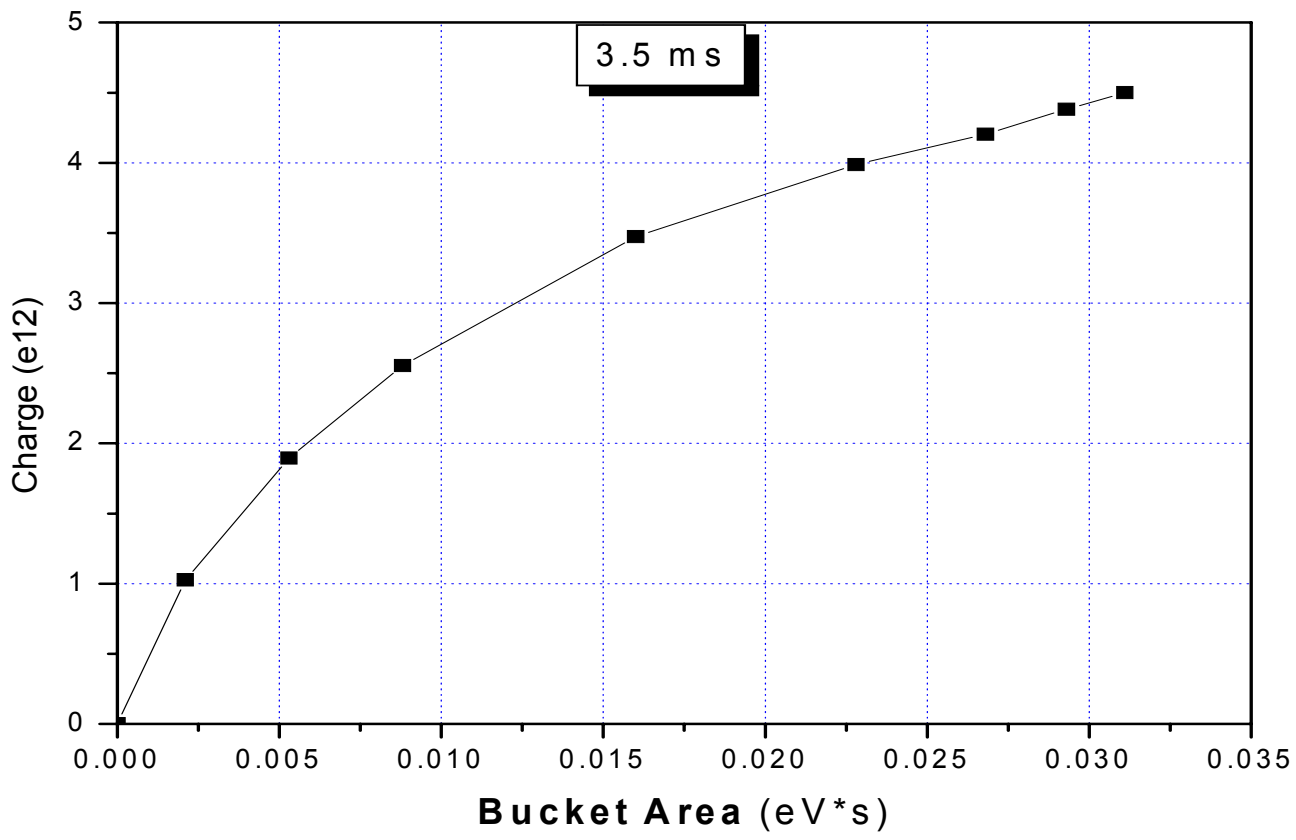

Fig. 5(e)

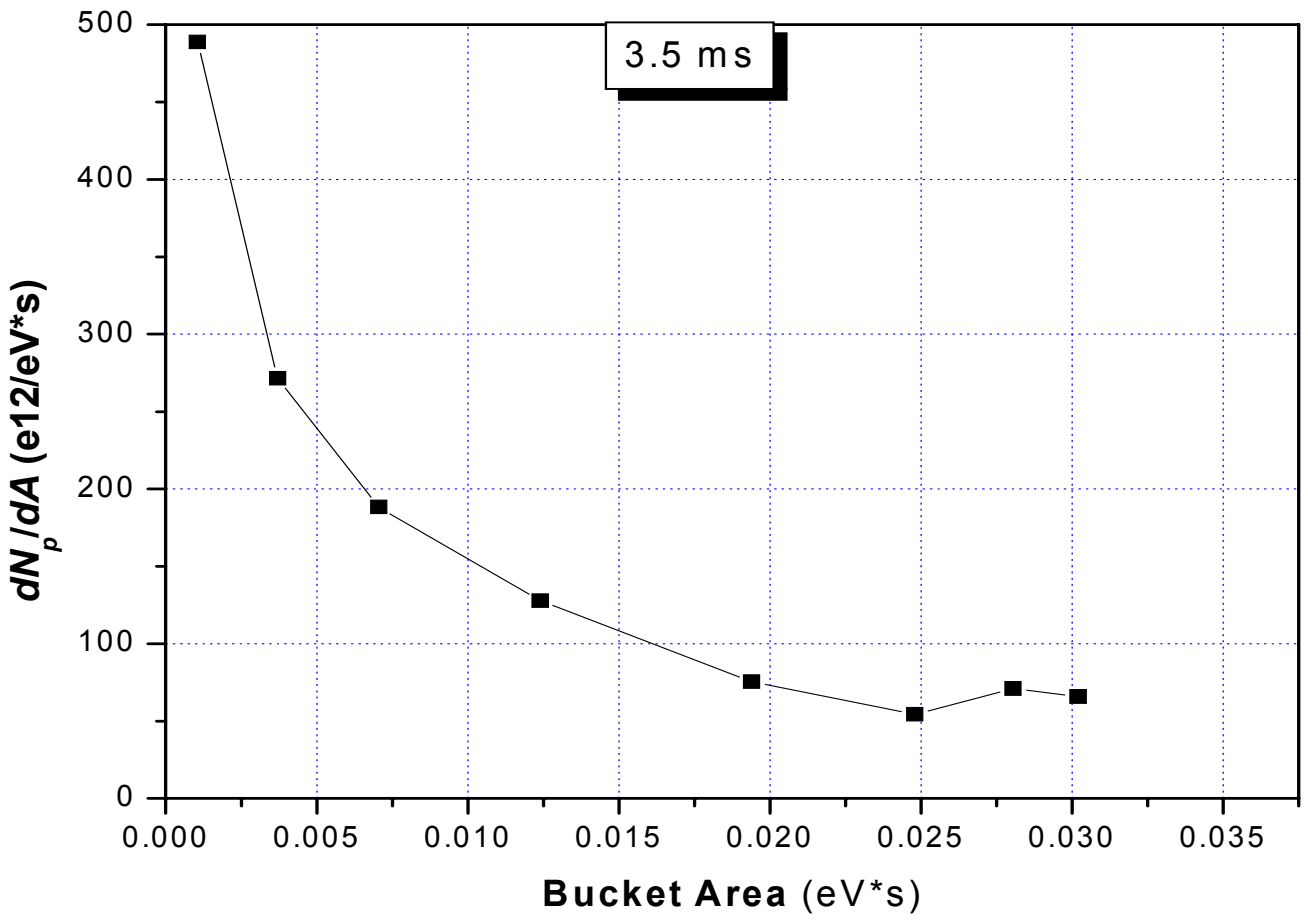

Fig. 5(f)

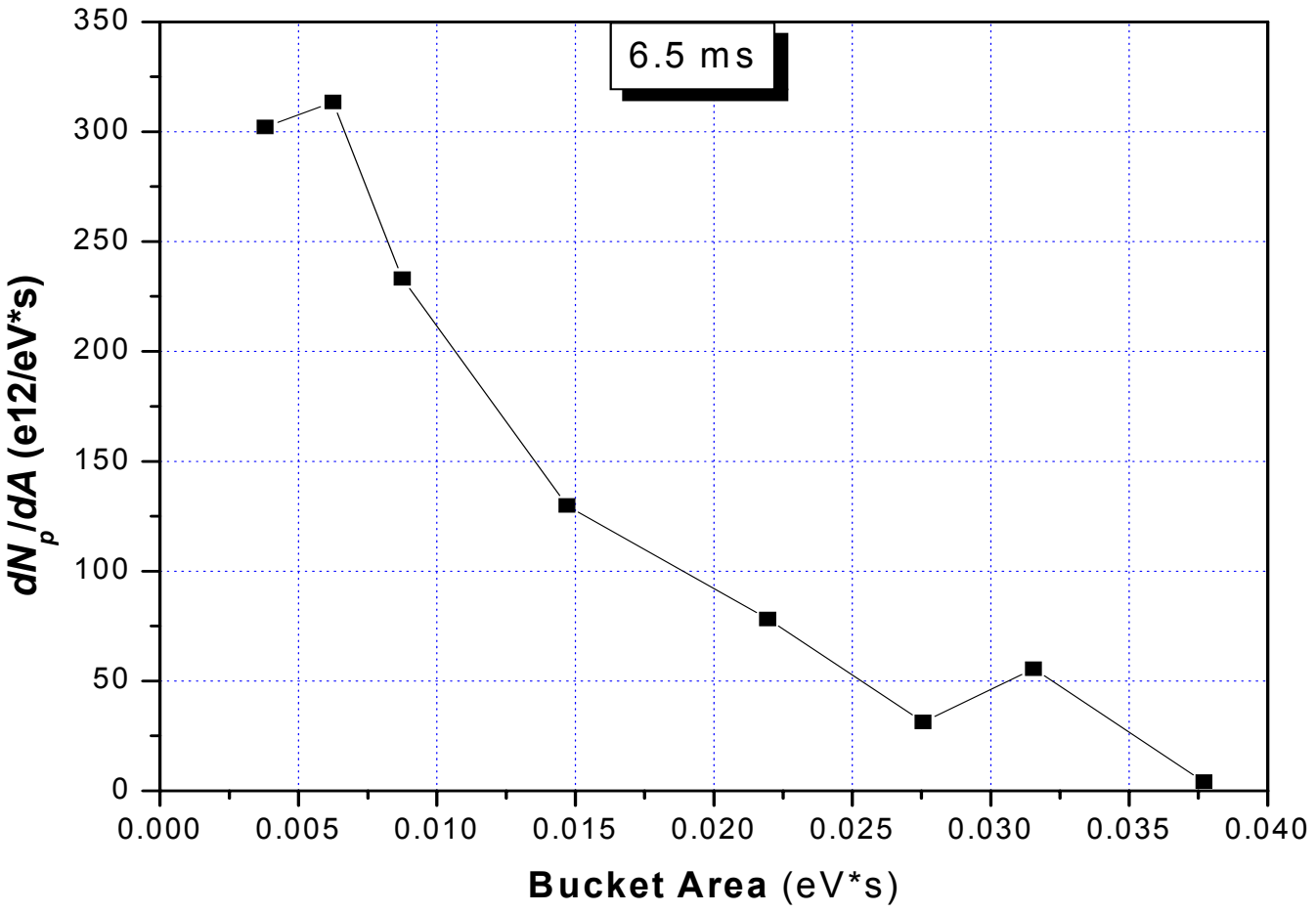

Fig. 5(g)

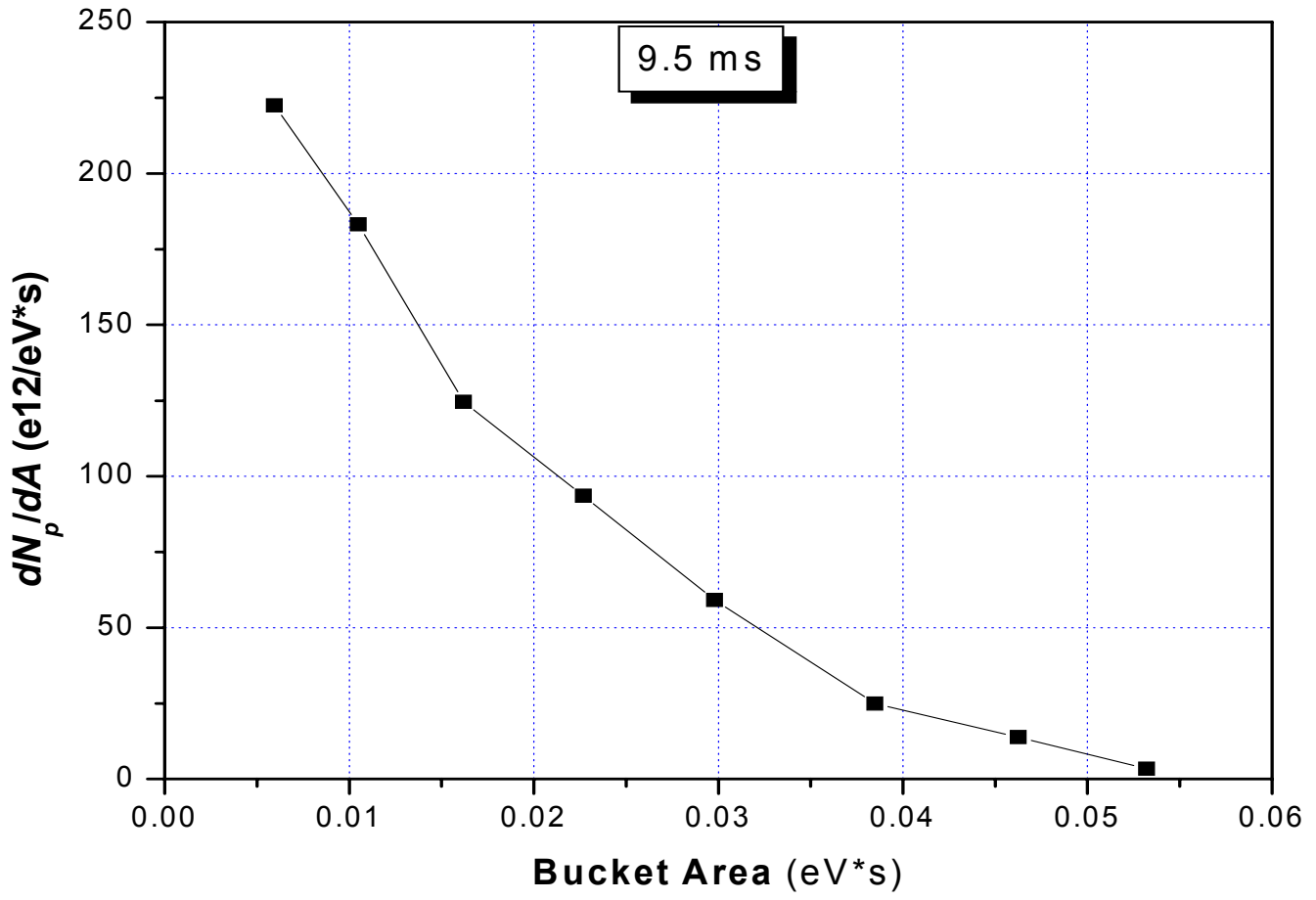

Fig. 5(h)

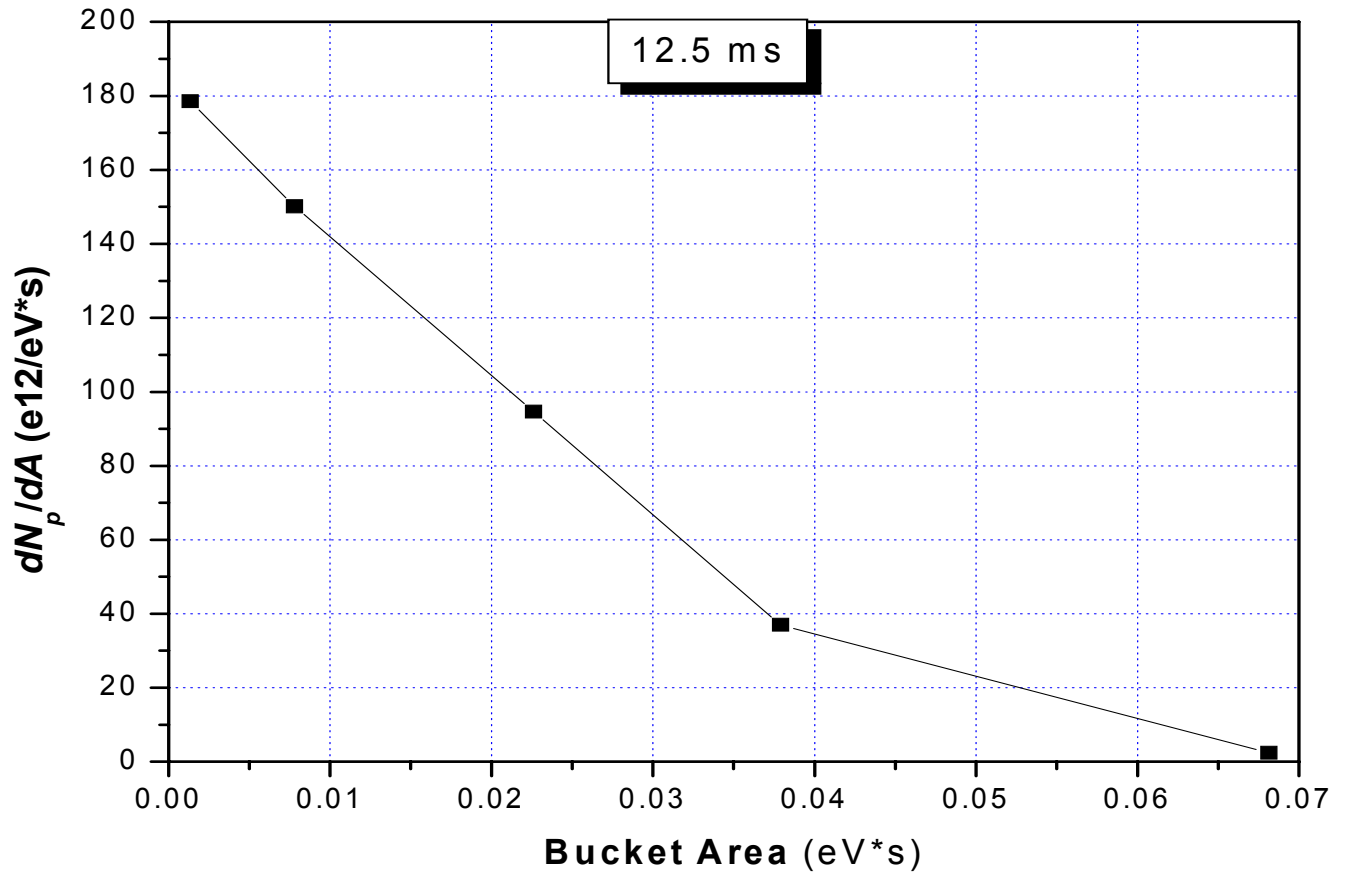

Fig. 5(i)

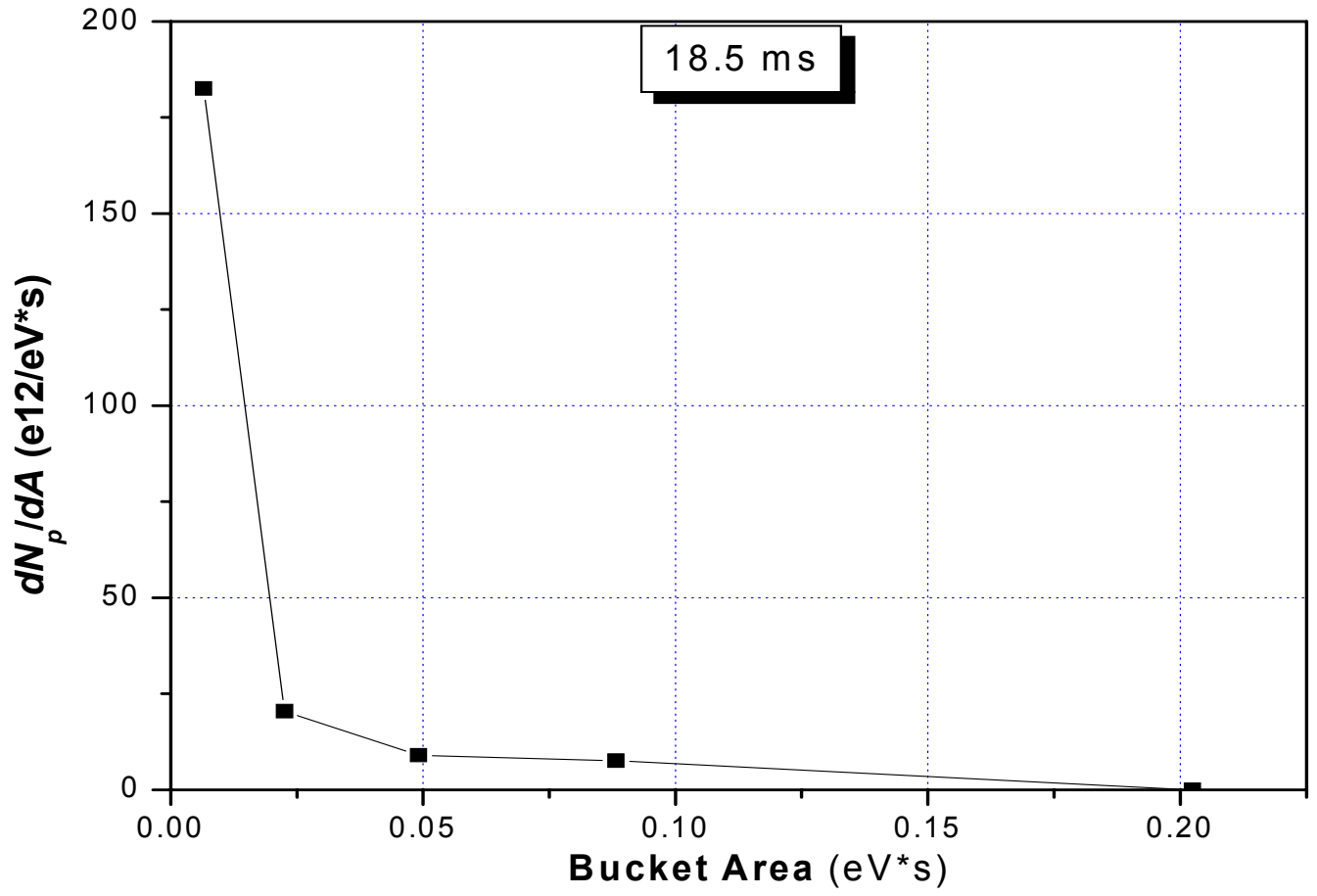

Fig. 5(j)

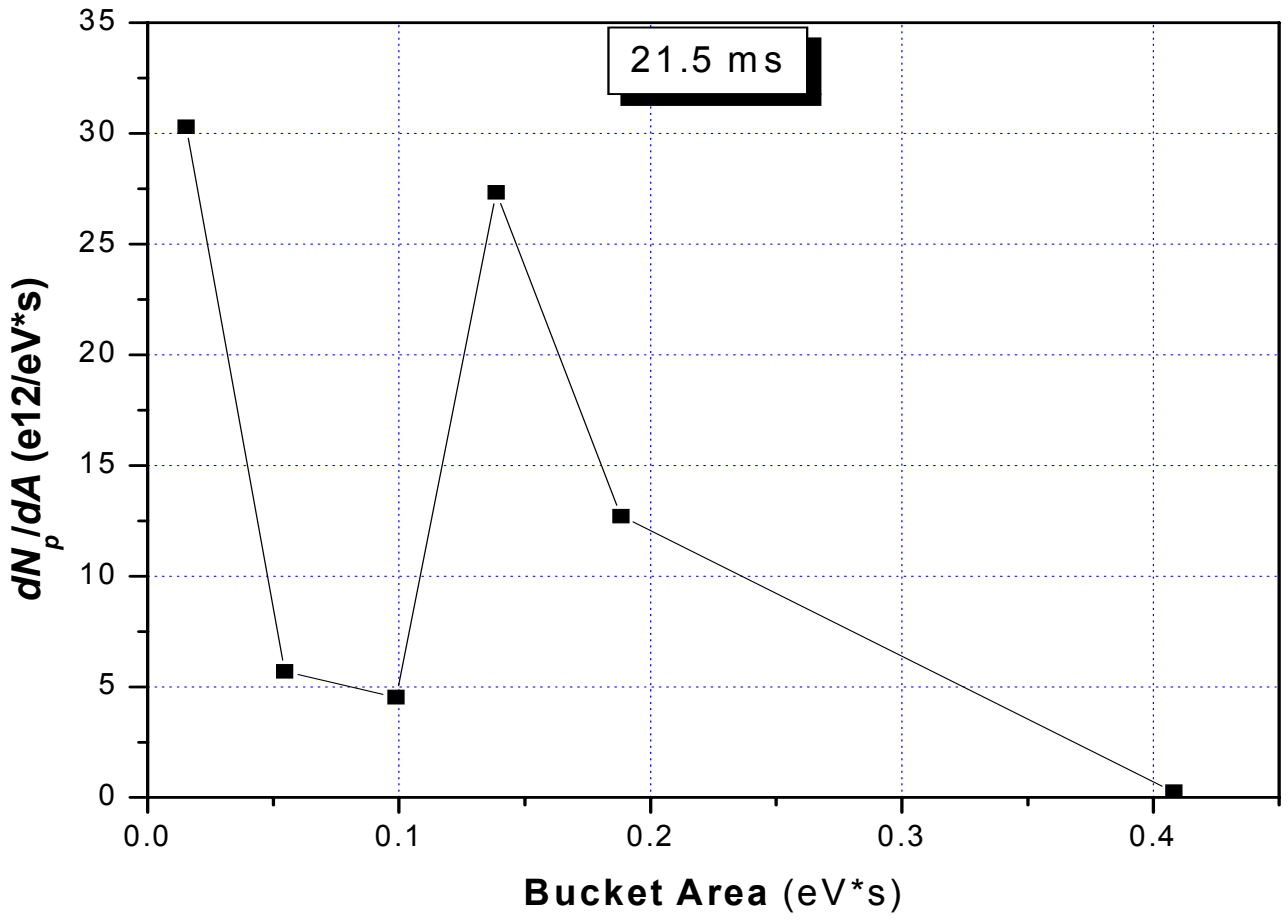

Fig. 5(k)

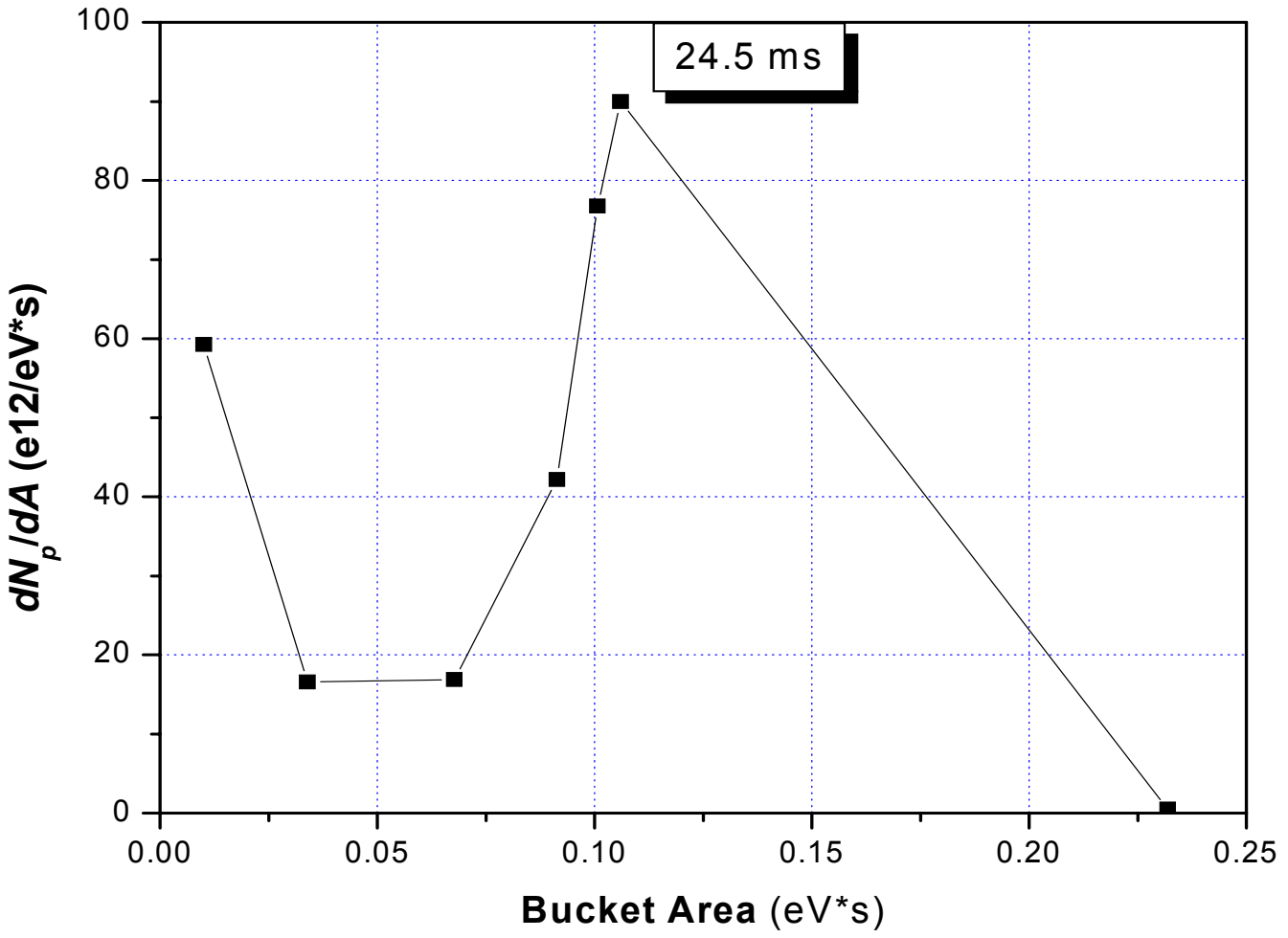

Fig. 5(l)

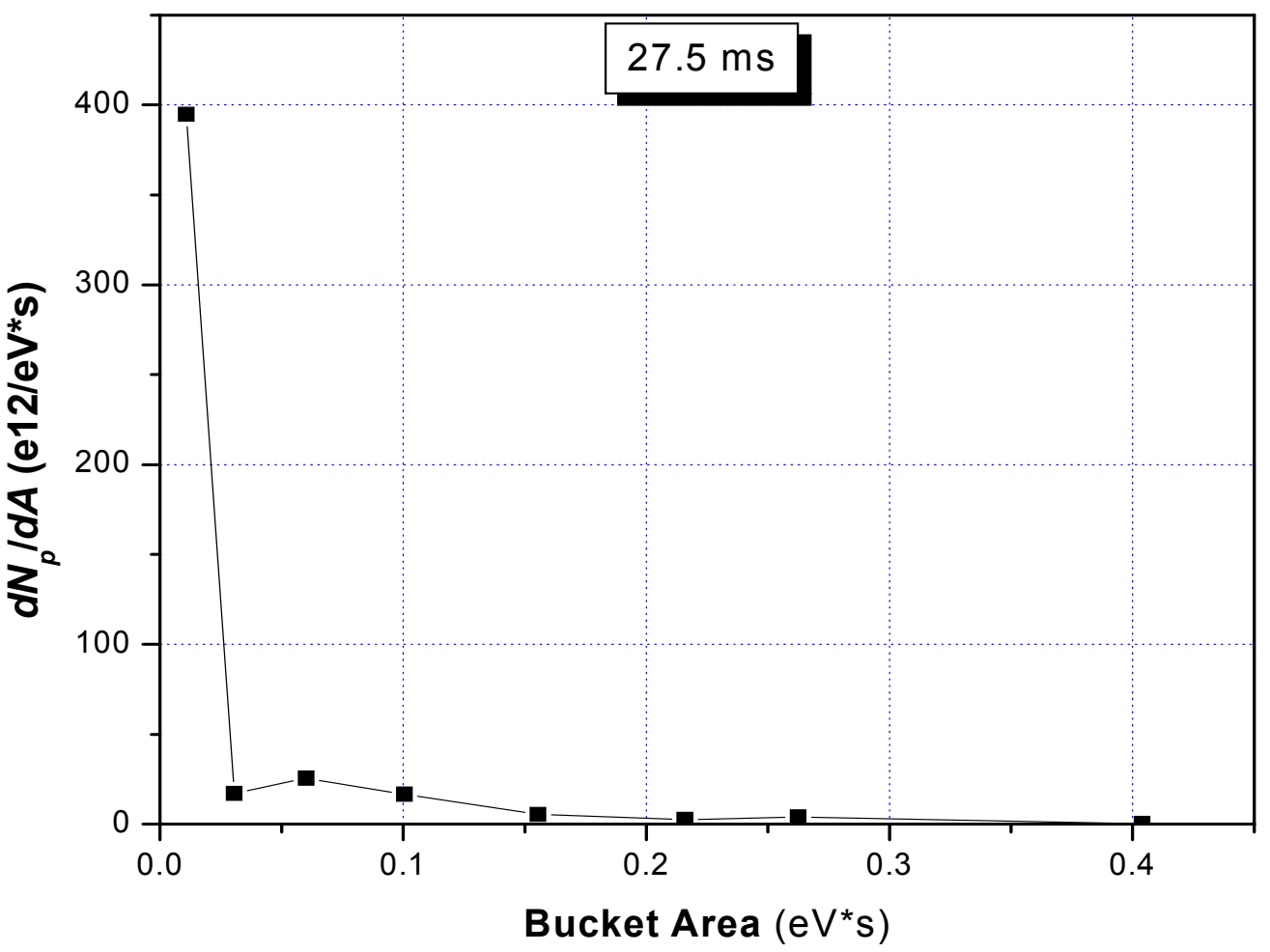

Fig. 5(m)

Fig. 5(a) at 3.5 ms, RFSUM was reduced to nine different values.

Fig. 5(b) the corresponding charge signal of Fig. 5(a).

Fig. 5(c) $\varphi_s$ *vs*. $V_{RF}$ at 3.5 ms.

Fig. 5(d) $V_{RF}$ and $\varphi_s$ *vs*. charge at 3.5 ms are shown as the black and blue curves respectively.

Fig. 5(e) the charge *vs*. bucket area at 3.5 ms.

Fig. 5(f) the charge density *vs*. bucket area at 3.5 ms.

Fig. 5(g) the charge density *vs*. bucket area at 6.5 ms.

Fig. 5(h) the charge density *vs*. bucket area at 9.5 ms.

Fig. 5(i) the charge density *vs*. bucket area at 12.5 ms.

Fig. 5(j) the charge density *vs*. bucket area at 18.5 ms.

Fig. 5(k) the charge density *vs*. bucket area at 21.5 ms.

Fig. 5(l) the charge density *vs*. bucket area at 24.5 ms.

Fig. 5(m) the charge density *vs*. bucket area at 27.5 ms.